# Ultrasmall $CsPbBr_3$ Blue Emissive Perovskite Quantum Dots using K-alloyed $Cs_4PbBr_6$ Nanocrystals as Precursors

*Clara Otero Martínez[1,2,‡], Matteo L. Zaffalon[3,‡], Yurii Ivanov[4], Nikolaos Livakas[2,5], Luca Goldoni[6], Giorgio Divitini[4], Sankalpa Bora[7], Gabriele Saleh[2], Francesco Meinardi[3], Andrea Fratelli[3], Sudip Chakraborty[7], Lakshminarayana Polavarapu[1], Sergio Brovelli[3]\*, Liberato Manna[2]\**

[1] CINBIO, Department of Physical Chemistry, Materials Chemistry and Physics Group, Universidade de Vigo, Campus Universitario As Lagoas-Marcosende, 36310 Vigo, Spain

[2] Nanochemistry, Istituto Italiano di Tecnología, Via Morego 30, 16163 Genova, Italy

[3] Dipartimento di Scienza dei Materiali, Università degli Studi di Milano-Bicocca, Via R. Cozzi 55, 20125, Milano, Italy

[4] Electron Microscopy and Nanoscopy, Istituto Italiano di Tecnología, Via Morego 30, 16163 Genova, Italy

[5] Dipartimento di Chimica e Chimica Industriale, Università di Genova,16146 Genova, Italy

[6] Material Characterization Facility, Istituto Italiano di Tecnologia Via Morego 30, 16163 Genova, Italy

[7] Harish-Chandra Research Institute (HRI) Allahabad, Jhunsi, Prayagraj, 211019, India

[‡] Equal Contribution

AUTHOR INFORMATION

**Corresponding Author**

*Liberato Manna - Nanochemistry, Istituto Italiano di Tecnologia, 16163 Genova, Italy; orcid.org/0000-0003-4386-7985; Email: liberato.manna@iit.it

*Sergio Brovelli - Dipartimento di Scienza dei Materiali, Università degli Studi di Milano-Bicocca, 20125 Milano, Italy; orcid.org/0000-0002-5993-855X; Email: sergio.brovelli@unimib.it

ABSTRACT

We report a colloidal synthesis of blue emissive, stable cube-shaped $CsPbBr_3$ quantum dots (QDs) in the strong quantum confinement regime *via* a dissolution-recrystallization starting from pre-synthesized $(K_xCs_{1-x})_4PbBr_6$ nanocrystals which are then reacted with $PbBr_2$. This is markedly different from the known case of $Cs_4PbBr_6$ nanocrystals that react within seconds with $PbBr_2$ and get transformed into much larger, green emitting $CsPbBr_3$ nanocrystals. Here, instead, the conversion of $(K_xCs_{1-x})_4PbBr_6$ nanocrystals to $CsPbBr_3$ QDs occurs in a time span of hours, and tuning of the QDs size is achieved by adjusting the concentration of precursors. The QDs exhibit excitonic features in optical absorption that are tunable in the 420 - 452 nm range, accompanied by blue photoluminescence with quantum yield around 60%. Detailed spectroscopic investigations in both the single and multi-exciton regime reveal the exciton fine structure and the effect of Auger recombination of these $CsPbBr_3$ QDs, confirming theoretical predictions for this system.

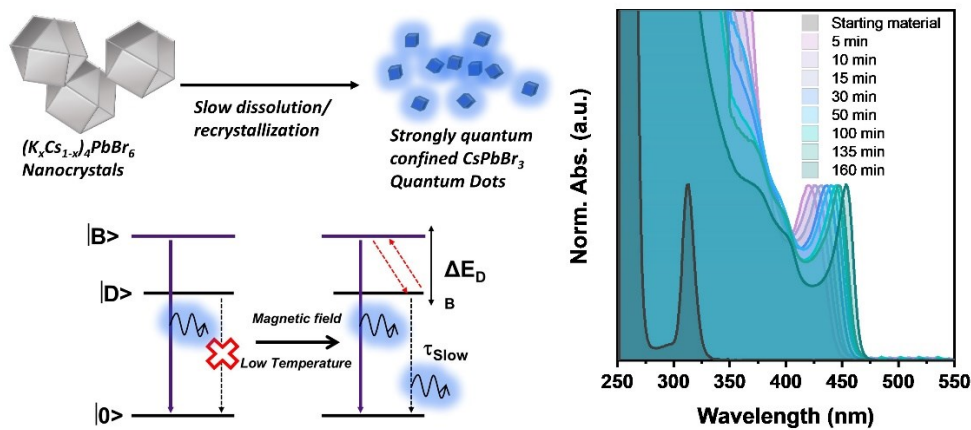

Colloidal lead halide perovskite nanocrystals, with $CsPbBr_3$ being the flagship stable material of the class, are a family of semiconductor materials with great potential for optoelectronic applications, such as light-emitting diodes (LEDs), quantum light emission, scintillators, and lasers, thanks to their excellent optical properties, which include narrow emission in the visible range of the spectrum with near unity photoluminescence quantum yield (PLQY).[1-3] In $CsPbBr_3$, the crystal structure is based on a 3D network of corner-sharing $[PbBr_6]^{4-}$ octahedra that are charge-balanced by $Cs^+$ cations positioned in the voids created by such network. Other Cs-Pb-Br phases are known, such as non-luminescent $Cs_4PbBr_6$ and $Cs_2PbBr_5$, the former being made of disconnected $[PbBr_6]^{4-}$ octahedra (hence a "0D" phase) separated by $Cs^+$ cations, and the latter made of layers of connected $[Pb_2Br_5]^-$ anions alternating with layers of $Cs^+$ cations (hence a "2D" phase).[4-9] It has been shown that the three phases can be interconverted, for example by $PbBr_2$ addition (0D → 3D; 3D → 2D)[6, 10, 11] or by CsBr extraction (0D → 3D; 2D → 3D).[12, 13] In this work, we show that in colloidal nanocrystals the 0D → 3D interconversion can be remarkably slowed down, such that, starting from 0D $Cs_4PbBr_6$ nanocrystals, these are slowly used up to synthesize ultrasmall $CsPbBr_3$ nanocrystals in the strong quantum confinement regime.

For colloidal semiconductor nanocrystals to be classified as "quantum dots" (QDs) their size, $d$, should be comparable to or smaller than the exciton Bohr diameter ($d_B$) of that material, which is around 7 nm for $CsPbBr_3$.[14] Conventional methods for the colloidal synthesis of $CsPbBr_3$ nanocrystals usually deliver relatively "large" $CsPbBr_3$ nanocrystals (with $d \sim 8$ nm or beyond)[14, 15] in a time range of seconds.[2] It is possible to grow smaller nanocrystals, for example by reducing the reaction temperature,[16-20] modulating the precursors ratios,[18] working in an environment with high concentration of ligands[19] or using "bulky" ligands with long alkyl chains,[20, 21] but only a few groups have succeeded to prepare stable $CsPbBr_3$ QDs with sizes below 5 nm.[18-20, 22-27] Dong et al. demonstrated the synthesis of $CsPbBr_3$ QDs below 5 nm size by tuning the Pb/Br ratio using $ZnBr_2$ as Br extra source.[18] More recently, Akkerman et al. were able to exert a tighter control on the nanocrystals growth kinetics by using tryoctylphosphine oxide (TOPO) and diisooctylphosphinic acid (DOPA) as weak coordinating ligands, achieving a much slower growth rate compared to synthesis schemes that use more conventional ligands such as oleylammonium (OLAm) and oleate (OL). They also managed to isolate $CsPbBr_3$ QDs smaller than 4 nm by stabilizing them with lecithin, a long chain zwitterionic ligand.[28] A similar approach was employed by Bi et al. to stabilize and isolate $CsPbBr_3$ QDs through a post-synthetic ligand exchange with phenylethylamine.[26] There have also been reports on the synthesis of ultra-small $CsPbBr_3$ nanocrystals with intermediate size between molecule and QDs, usually named "nanoclusters".[29-32] However, over time they tend to aggregate. These clusters have mainly been used as "precursors" to prepare $CsPbBr_3$ nanocrystals of lager sizes, different geometries[31, 33-35] or even to grow perovskite-based heterostructures.[34, 36-40]

Another way of slowing down the nucleation and growth of $CsPbBr_3$ nanocrystals was identified by Su et al.[41] through the addition of $Ag^+$ ions. Monovalent cations smaller than $Cs^+$ (such as $Ag^+$ or alkali metals) cannot penetrate into the crystal lattice in significant amounts to form alloy compositions, and at best they can enter in small amounts as dopants.[42, 43] On the other hand, alkali metal cations such as $K^+$ and $Na^+$ can bind to the surface of perovskite nanocrystals (most likely by forming stable metal-halide bonds).[44-46] The presence of surface-bound alkali metal cations may suppresses the formation of trap states and consequently the ion migration, but can also prevent the aggregation/coalescence of nanocrystals. For example, platelet-shaped $CsPbBr_3$ nanocrystals were better able to preserve their shape and optical characteristics under photo and thermal stress over time when treated with $K^+$ cations, compared to untreated platelets.[45]

Here, we capitalize on the various previous findings, as highlighted above, and report a new route to synthesize stable, blue emissive $CsPbBr_3$ QDs in the strong quantum confinement regime, starting from 0D $(K_xCs_{1-x})_4PbBr_6$ nanocrystals that were used as reactants. A first key discovery of our work is that $Cs_4PbBr_6$ nanocrystals can be alloyed with a substantial amount of $K^+$ ions, reaching a close to 20% replacement of $Cs^+$ with $K^+$ ions (**Figure 1**). This is markedly different from the case of $CsPbBr_3$ discussed above, in which alloying with $K^+$ is not observed. Crucial to this work, these $(K_xCs_{1-x})_4PbBr_6$ nanocrystals are much less reactive than the unalloyed $Cs_4PbBr_6$ ones towards $PbBr_2$. The addition of $PbBr_2$ to $Cs_4PbBr_6$ nanocrystals is known to rapidly transform them to $CsPbBr_3$ nanocrystals,[6] whereas the $(K_xCs_{1-x})_4PbBr_6$ nanocrystals are much less reactive and follow a dissolution-recrystallization reaction pathway: they essentially act as reservoirs, steadily getting dismantled and releasing monomers for the slow nucleation and growth of $CsPbBr_3$ QDs in a size range between $d$=3 nm and $d$=3.5 nm. These QDs, once isolated and purified, remain stable and do not evolve further in size nor they aggregate. This is markedly different from previous works on the synthesis of QDs which required the addition of bulky ligands (such as lecithin[28] or phenylethylamine[26]) to preserve colloidal stability once the QD were isolated from the reaction environment. At present, the reasons for the increased stability of our QDs are unclear and we can only speculate that it might stem from the presence of trace amounts of $K^+$ ions on the QD surface, although this remains a hypothesis, as the concentration of $K^+$ ions in the final sample is below the experimental error.

The much higher stability of these QDs compared to previously reported materials in the same $d$-range enabled a more careful study of the photophysics of $CsPbBr_3$ in the strong quantum confinement regime. We therefore investigated both the single and multi-excitonic behaviour for QDs with $d$~3.5 nm. Cryogenic time-resolved photoluminescence (PL) experiments, corroborated for the first time for halide perovskite QDs by fluorescence line-narrowing (FLN) measurements in both continuous wave (*cw*) and time-resolved modes, confirm the characteristic signatures of a dark exciton state separated from the overlying bright exciton state by a dark-bright splitting energy $\Delta_{DB}$=15 meV for $d$=3.5 nm, in full agreement with the theoretical predictions of Efros and co-workers[47], and extending the size range for which the exciton structure is now known. Also consistently, control samples of the tiniest QDs we could synthesize by conventional hot injection methods ($d$=4 nm) featured a smaller $\Delta_{DB}$ (10.4 meV), and growing larger particles further modified the exciton fine structure, leading to a level crossover driven by the Rashba effect, resulting in a lower lying bright exciton state and accelerated radiative decay at low temperatures. We also performed transient absorption (TA) measurements to assess the influence of non-radiative Auger recombination on the biexciton yield in strongly confined QDs. The biexciton lifetime for $d$=3.5 nm QDs is estimated in 8.2 ps, which agrees well with the projected volume-scaling and sets a new small-size limit to our estimation of the Auger process in $CsPbBr_3$ QDs.

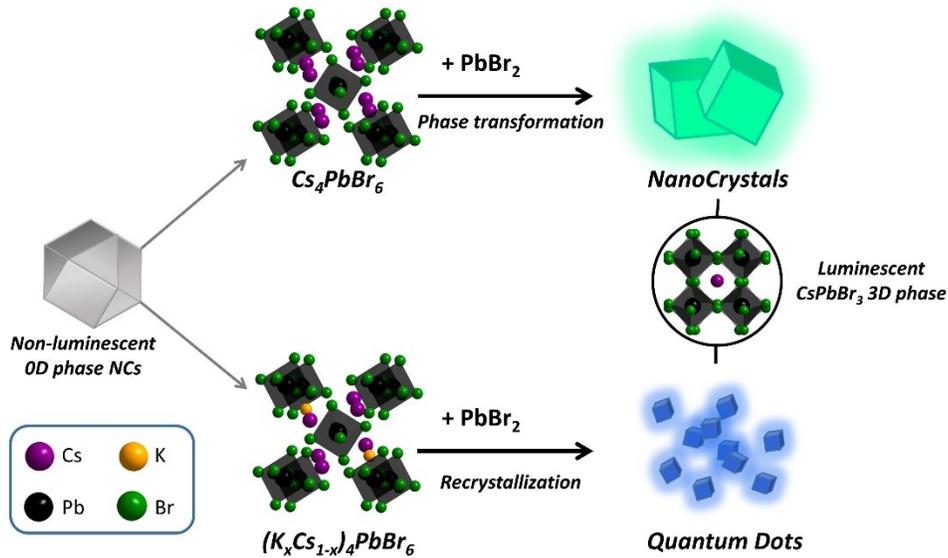

**Figure 1.** Sketch of the conversion of 0D-phase $Cs_4PbBr_6$ nanocrystals to 3D-phase $CsPbBr_3$ nanocrystals/QDs by $PbBr_2$ addition. Top: unalloyed $Cs_4PbBr_6$ nanocrystals yield large $CsPbBr_3$ nanocrystals through a fast reaction with $PbBr_2$, while K-alloyed $(K_xCs_{1-x})_4PbBr_6$ nanocrystals yield < 3.5 nm size $CsPbBr_3$ QDs through a slow dissolution-recrystallization process.

The starting $Cs_4PbBr_6$ nanocrystals were synthesized following a previously reported hot injection method (see experimental section in SI for more details).[6] The crude nanocrystals solution was purified by centrifugation (without using any antisolvent) and the precipitated nanocrystals were then redispersed in toluene. The K-alloyed $(K_xCs_{1-x})_4PbBr_6$ nanocrystals were obtained from this sample by a partial cation exchange reaction (**Figure 2a**) which was performed by adding a solution of potassium oleate to the colloidal $Cs_4PbBr_6$ nanocrystals dispersion, and letting the mixture react for a few seconds. This was followed by the addition of ethyl acetate (EtOAc) as antisolvent, precipitation of the nanocrystals via centrifugation, and their redispersion in a solution of oleylamine and oleic acid in toluene, to compensate for the partial loss of ligands that is likely to occur during the purification. The procedure was repeated three times to guarantee the maximum incorporation of potassium. The UV-Vis absorption spectra of the nanocrystals before and after the reaction with K-oleate are shown in **Figure 2b**. After the reaction, the characteristic excitonic peak at ~ 314 nm corresponding to the $Cs_4PbBr_6$ nanocrystals was shifted to slightly higher energies (~ 311 nm). Neither of the samples emits light. Based on transmission electron microscopy (TEM) and High Annular Dark Field (HAADF)-Scanning TEM, the hexagonal shape (in projection) of the starting $Cs_4PbBr_6$ nanocrystals was preserved after the partial cation exchange reaction (**Figure 2c-f, S1 & S2**), and their average size slightly decreased from 17.8 to 17.0 nm (see also **Figure S1 & S2**), caused by a lattice contraction by the introduction of potassium, due to the smaller ionic radius of $K^+$ with respect to $Cs^+$ (1.38 vs 1.68 pm),[48] and also some etching of the nanocrystals, due to their exposure to a mild polarity during the several cycles of washing with EtOAc.

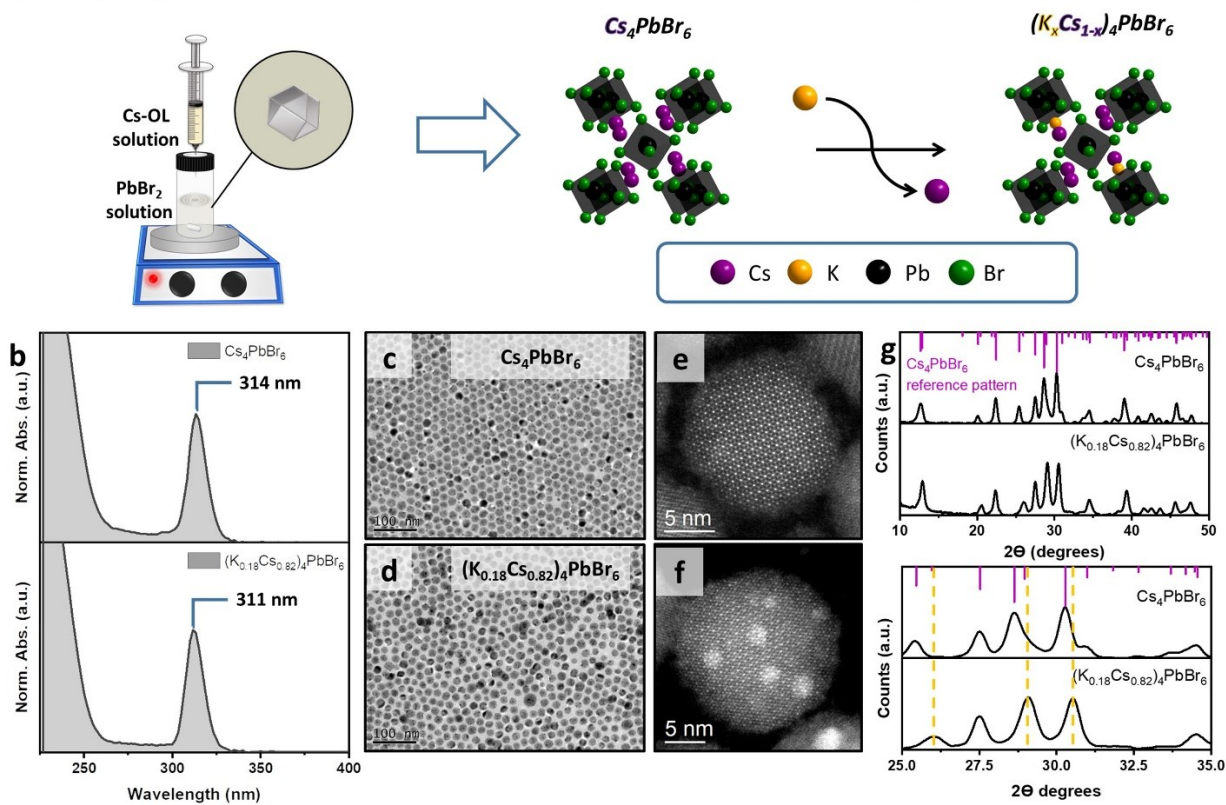

**Figure 2.** (a) Schematic illustration of the synthesis of $Cs_4PbBr_6$ nanocrystals by the hot injection of a Cs-oleate (Cs-OL) solution into a vial containing the $PbBr_2$ solution. (b-.g): UV-Vis absorption spectra (b), low magnification TEM images (c-d), corresponding high resolution HAADF-STEM images at the [001] zone axis (e-f), and XRD diffraction patterns (g) of $Cs_4PbBr_6$ nanocrystals and the corresponding $(K_xCs_{1-x})_4PbBr_6$ nanocrystals obtained by partial $Cs^+ \rightarrow K^+$ cation exchange. The scale bar in panels (c-d) is 100 nm. In panel (g) bottom, the XRD pattern in the region between 25° and 35° highlights the peak shifts in the alloyed sample compared to the initial sample.

The X-ray diffraction (XRD) pattern of the pristine sample (**Figure 2g**) matched with the typical features of the trigonal $R\bar{3}c$ space group, and was in agreement with previous reports.[49] After the reaction with $K^+$, the XRD peaks were shifted to higher angles (see the zoomed diffraction pattern in the bottom panel of **Figure 2g**). This is ascribed to the contraction of the cell due to the partial replacement of $Cs^+$ with $K^+$ (**Figure 2a & g**). The calculated lattice parameters of the pristine and final alloy samples (**Figure S3**) confirm a 1.1% decrease in the unit cell volume following the exchange reaction. The elemental composition of the two samples was determined by scanning electron microscopy – energy dispersive X-ray spectroscopy (SEM-EDS) and by HAADF-STEM-EDS (see **Figures S4-S7**). According to the analyses, the pristine sample matches a $Cs_4PbBr_6$ stoichiometry and the final alloy sample matches a $(K_{0.18}Cs_{0.82})_4PbBr_6$ one. Hence the $Cs^+ \rightarrow K^+$ exchange was not complete and most likely reached a thermodynamic equilibrium. The difficulty in achieving a more extensive exchange is also evident from the fact that there is no known stable $K_4PbBr_6$ phase.

We performed density functional theory (DFT) calculations to elucidate the effect of $K^+$ alloying on the electronic structure of $Cs_4PbBr_6$ (**Figure S8**). We systematically explored all the possible $K^+$ substitution positions in order to find the lowest-energy configuration. Due to the resulting combinatorial complexity, the simulated doping concentration was limited to 12.5%, which is however sufficient to study the effect of $K^+$ alloying. Upon $K^+$ alloying, the band gap increases from 3.91 to 3.96 eV, in good agreement with experimental findings (**Figure 2b**). The inspection of the density of states reveals that neither $Cs^+$ nor $K^+$ participates to the electronic states near the band edges (**Figure S9**). The slight band gap widening upon $K^+$ doping can thus be ascribed to the resulting lattice contraction (-2.2% in volume according to DFT). Both the valence and conduction bands display a dominant contribution from Pb orbitals and a smaller but significant contribution from Br orbitals. The comparison with 3D lead halide perovskites suggests that the valence and conduction bands are formed by Br(p)-Pb(s) and Br(p)-Pb(p) antibonding states, respectively.[50, 51] However, unlike 3D perovskites, in 0D perovskites the corresponding bands are quite flat (low dispersion), since they localize in isolated $PbBr_6$ octahedra (**Figure S10)**. This also indicates a much higher electron and hole effective masses in 0D perovskites. Finally, we note that in both doped and undoped $Cs_4PbBr_6$ the band gap is indirect (**Figure S10**), possibly explaining the absence of photoluminescence in these compounds.

We carried out analyses via nuclear magnetic resonance (NMR) spectroscopy to characterize the ligand shell of the two systems (the non-doped $Cs_4PbBr_6$ nanocrystals and the $(K_xCs_{1-x})_4PbBr_6$ ones). Details are reported in the SI (**Figures S11-14**). These analyses evidenced a marked difference in the composition and density (**SEq. 1-2**) of the ligand shell for the two systems. The $Cs_4PbBr_6$ nanocrystals were characterized by a densely packed ligand shell (~ 3.5 ligands/nm$^2$) constituted by 78% oleate and 22% oleyl ammonium. On the other hand, the $(K_xCs_{1-x})_4PbBr_6$ nanocrystals were characterized by a significantly less packed ligand shell (~2.3 ligands/nm$^2$) again mainly made of oleate (81%), with only 19% oleylammonium. These marked differences in the ligand shell density for the two systems are probably responsible for their much different reactivities toward $PbBr_2$, as discussed in the following section.

The reactivity of the synthesized $Cs_4PbBr_6$ and $(K_{0.18}Cs_{0.82})_4PbBr_6$ nanocrystals in toluene towards a $PbBr_2$ solution was investigated regarding their chemical and morphological transformations. The $Cs_4PbBr_6$ nanocrystals were already reactive at room temperature and quickly converted to 12.2 nm cube-shaped $CsPbBr_3$ nanocrystals having a strong green emission (**Figure S15**), as already shown by our group in a previous work.[6] The $(K_{0.18}Cs_{0.82})_4PbBr_6$ nanocrystals were instead much less reactive. Mixing the $(K_{0.18}Cs_{0.82})_4PbBr_6$ nanocrystals solution with the $PbBr_2$ solution at room temperature only produced an absorption spectrum that was the sum of the two components (**Figure 3a**, second panel from the top, **& Figure 3b**).

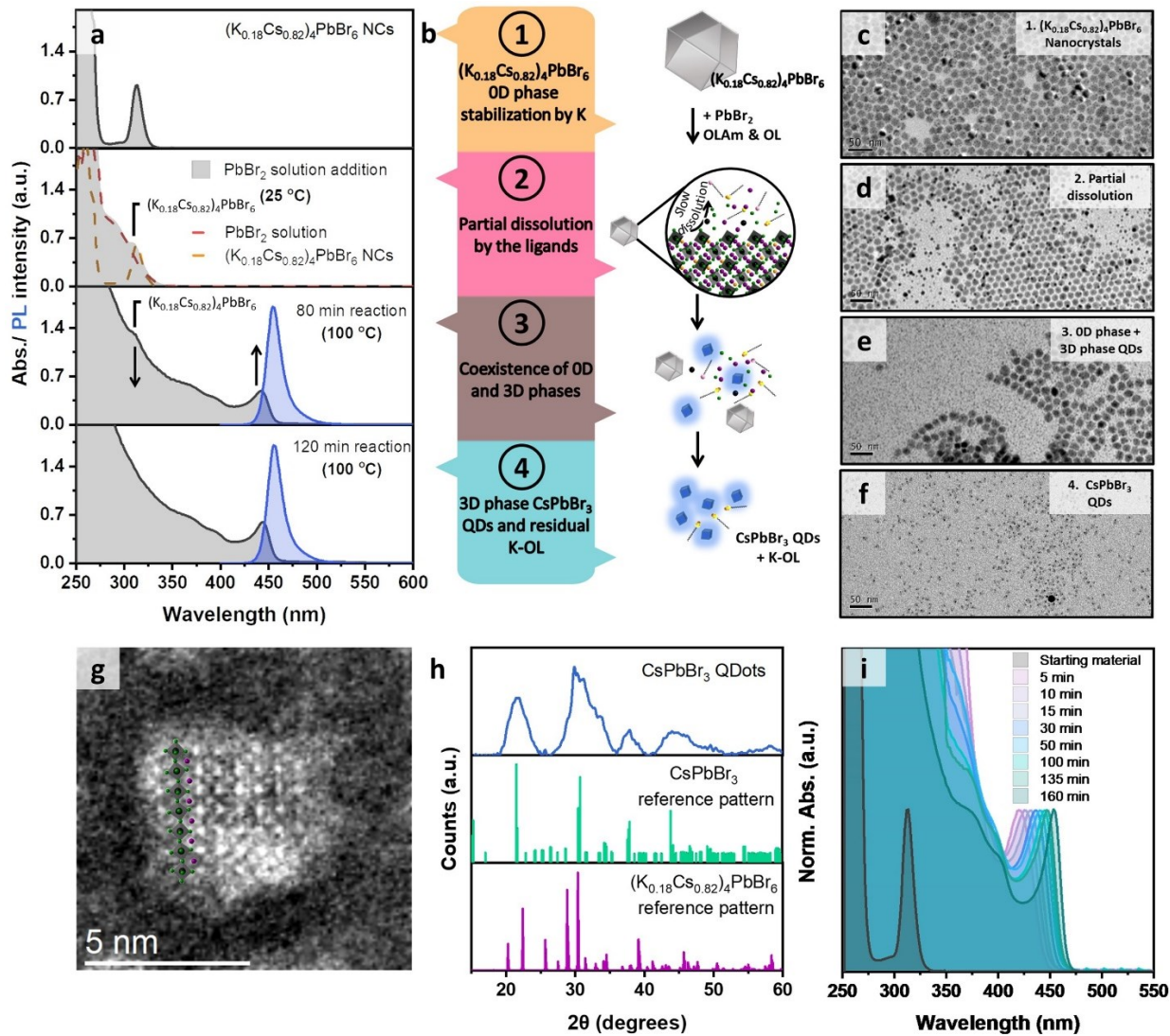

**Figure 3.** (a) UV-Vis absorption and PL spectra of the 0D phase $(K_xCs_{1-x})_4PbBr_6$ conversion to 3D phase $CsPbBr_3$ QDs at different reaction stages. (b) Schematic illustration of the proposed reaction mechanism. (c-f) TEM images of the 0D phase $(K_xCs_{1-x})_4PbBr_6$ conversion to 3D phase $CsPbBr_3$ QDs at different reaction stages. (g) High resolution HAADF-STEM image of a representative QD sample. The images were acquired using a low electron dose to avoid the degradation of the particles due to the beam sensitivity of the material. The dimensions of the inorganic core along one edge of the cube correspond to about 6 unit cells (~ 3.5 nm), based on the number of atom columns with higher contrast (Pb-Br columns). (h) XRD diffraction pattern of the 3.5 nm $CsPbBr_3$ QDs (top panel) and comparison to $(K_xCs_{1-x})_4PbBr_6$ and $CsPbBr_3$ reference patterns. (i) UV-Vis absorption spectra of $CsPbBr_3$ QDs of different sizes obtained at 100 °C using different reaction times.

To observe the evolution in the new optical features, the solution had to be heated to 100 °C, after which the optical absorption features typical of very small CsPbBr$_3$ nanocrystals started to emerge. This can be seen in **Figure S16a**, which is a zoom in of the blue region of the spectra. At the same time, an emission feature in the green region of the spectrum was seen, indicating the concomitant presence of a small fraction of larger nanocrystals, which however almost completely disappeared at later reaction times (**Figure S16b**). It is important to note that, while these features emerged and evolved, we could still monitor the persistence in the optical absorption spectrum of the excitonic band at ~ 311 nm assigned to the (K$_{0.18}$Cs$_{0.82}$)$_4$PbBr$_6$ nanocrystals, which however steadily decreased in intensity. This indicates the coexistence of the two materials: the initial (K$_{0.18}$Cs$_{0.82}$)$_4$PbBr$_6$ nanocrystals, steadily being consumed, and the newly formed CsPbBr$_3$ QDs (**Figure 3a**, 3$^{rd}$ panel from the top **& Figure 3b**). It was not until after 2 hours of reaction that the signature of the (K$_{0.18}$Cs$_{0.82}$)$_4$PbBr$_6$ nanocrystals disappeared from the absorption spectrum (**Figure S17 & 3a**, 4$^{th}$ panel from the top). At this point, the absorption and PL spectra evidenced the presence of a monodisperse population of CsPbBr$_3$ QDs with a sharp excitonic band at ~445 nm and narrow blue PL emission peak at 455 nm with full width at half maximum of 15 nm (**Figure 3a**, 4$^{th}$ panel).

This and the different stages of the reaction depicted in **Figure 3b** were also corroborated by TEM analysis of the initial, intermediate, and final samples (**Figure 3c-f**) and the corresponding XRD pattern of the final product (**Figure 3h**). In XRD, the intermediate samples evidenced the presence of both the 0D and 3D phases (**Figure S18**), whereas in the final sample only the 3D phase was present (**Figure 3h**). Importantly, the broadening of the XRD peaks due to restricted number of reflecting planes corroborates the small size of these QDs. After 120 minutes, the reaction was completed and quenched in an ice-water bath. The solution had yellow colour and was colloidally stable. Due to the high concentration of ligands, the final sample was easily purified by their precipitation with antisolvents and was then redispersed in toluene. The peak positions in the XRD pattern of this final sample are fully in line with the pure $CsPbBr_3$ perovskite phase, indicating that the QDs do not incorporate any meaningful amount of $K^+$. This was also corroborated by elemental analysis via energy dispersive X-ray spectroscopy in the scanning electron microscope (SEM-EDS), which did not reveal any K trace. Yet, we cannot exclude a small, residual amount of $K^+$ ions on the surface of the QDs. An atomic resolution HAADF-STEM image of a representative QD from this sample is reported in **Figure 3g**.

The sluggish reactivity of the $(K_{0.18}Cs_{0.82})_4PbBr_6$ nanocrystals towards $PbBr_2$ compared to the pristine $Cs_4PbBr_6$ nanocrystals case suggests that the incorporation of $K^+$ ions in the 0D phase decreases their reactivity. This in turn can be ascribed to a different binding strength of the ligands, a different packing density of ligands, or yet to a decreased propensity of $PbBr_2$ species to intercalate in the $(K_{0.18}Cs_{0.82})_4PbBr_6$ alloy phase, due to the overall decrease in lattice parameter. We carried out further studies on the 0D → 3D conversion synthesis for better understanding of the mechanism. The time it takes for the $(K_{0.18}Cs_{0.82})_4PbBr_6$ nanocrystals to be completely depleted to nucleate and simultaneously grow into $CsPbBr_3$ QDs should likely depend on their initial concentration. Thus, in principle, the higher the concentration of these initial nanocrystals (hence their availability in the reaction medium), the longer it will take for them to be consumed and the larger will be the size of the final $CsPbBr_3$ QDs at that stage. To validate this hypothesis, several syntheses were performed using different volumes of $(K_{0.18}Cs_{0.82})_4PbBr_6$ nanocrystals precursor, while keeping all the other reaction parameters constant (see **Table S1** and experimental section in SI for more details). We indeed observed that the higher the volumes of $(K_{0.18}Cs_{0.82})_4PbBr_6$ nanocrystals, the longer the reaction times required to consume all the precursor material. The UV-Vis absorption spectra of the as-synthesized QDs (**Figure 3i**) exhibit sharp and tunable excitonic peaks from ~ 420 to 452 nm (**Figure 3i**) for the different reaction times. The samples exhibit emission peaks ranging from ~ 444 to ~ 462 nm, with full width at half maximum (FWHM) between 24 and 15 nm, respectively (**Figure S19**), and PLQY values in the range of 60 – 65 %. Only for short reaction times we could observe emission features in the green, stemming from a contamination of larger $CsPbBr_3$ nanocrystals, as mentioned earlier. This feature however disappeared for reaction times longer than 30 min. However, for reaction times longer than 160 min, the high concentration of $CsPbBr_3$ QDs in solution, along with the depletion of precursors (all the 0D nanocrystals were consumed at this time) initiated an Ostwald ripening process that eventually led to the formation of large green emissive $CsPbBr_3$ nanocrystals. Consistent with the size-independent PLQY values of the QD, the fluorescence dynamics was found to be essentially constant with increasing $d$ (**Figure S20**), with mono/bi-exponential decays with lifetimes around 4.5 ns. As a note, our method allows a relatively narrow range of tunability in the optical features (420-452 nm, see **Figure 3i**) that indicates a small tunability in size of the QDs in a size range peaked around 3-3.5 nm.

The stability of a sample of $CsPbBr_3$ QDs was monitored by UV-Vis absorption and PL spectroscopy for more than 1 year (**Figure S21**) in a closed vial under air. After this time, the QDs still exhibited strong blue emission (**Figure S21, inset**), although in the meanwhile the PL peak had shifted from ~ 455 nm to ~465 nm. The absorption spectra as well had shifted by a comparable amount, while the absorbance intensity was barely affected.

CsPbBr$_3$ QDs with $d < 4$ nm have been less explored than larger CsPbBr$_3$ nanocrystals. Currently, fine exciton structure studies for particles in the strong confinement regime are very scarce[24, 52-54] and so are investigations of the multiexcitonic recombination.[55] To fill this gap, we thus interrogated both the single and multiexciton photophysics by time-resolved PL as a function of temperature also in fluorescence line narrowing (FLN) configuration and in the presence of strong magnetic fields, and by TA experiments in the single vs. multiexciton regime. The PL decay curves of CsPbBr$_3$ QDs synthesized with our new approach, with $d = 3.5$ nm (**Figure 3g**), at decreasing temperature (300-2 K) are depicted in **Figure 4a**. The corresponding time and spectrally resolved PL decays at room temperature are reported in **Figure S22** showing uniform emission decay across the whole PL spectrum. Consistent with previous reports,[47, 56-58] cooling the CsPbBr$_3$ QDs from 300 K to 2 K progressively slowed down the PL dynamics, turning from an essentially single exponential behavior to a strongly biexponential kinetics dominated by a microsecond slow recombination tail. To better monitor the PL dynamics with temperature, we extracted the PL decay rates ($k$) and reported them in the inset of **Figure 4a** showing the gradual lengthening of PL decay which saturated at $k = 0.7$ μs$^{-1}$, corresponding to a PL lifetime τ~1.4 μs below 15 K. Crucially, this occurred at constant PLQY (**Figure S23**), indicating that the optical properties emerged from the band-edge excitonic fine structure comprising a low-lying dark excitonic state (|D>) with total angular momentum J=0 energetically separated by an optically active bright triplet state (|B>) with J=1, as theoretically predicted using electron-hole exchange[47, 59] or atomistic arguments[60] in the strong confinement regime. By assuming a Boltzmann distribution of excitons between the |D> and |B> states, we could model the transition between the slow and fast kinetics in a three-level scheme, by expressing the temperature dependence of the total decay rate as:

$$k(T) = \left(k_D + k_B \cdot e^{-\Delta_{DB}/\gamma T}\right)\left(1 + e^{-\Delta_{DB}/\gamma T}\right) \qquad \textit{Eq. 1}$$

where $k_D$ and $k_B$ are the radiative decay rates of the |D> and |B> states respectively, $\gamma$ is the Boltzmann constant, and $\Delta_{DB}$ being the dark-bright energy splitting[61]. The model well reproduced the experimental data, returning a splitting energy $\Delta_{DB}$ = 14.5 meV, in good agreement with the theoretical prediction for CsPbBr$_3$ QDs of this size[62, 63] and with experimental values measured for other strongly confined nanocrystals[64, 65] also considering possible differences in the dielectric confinement term of the samples analyzed in literature.

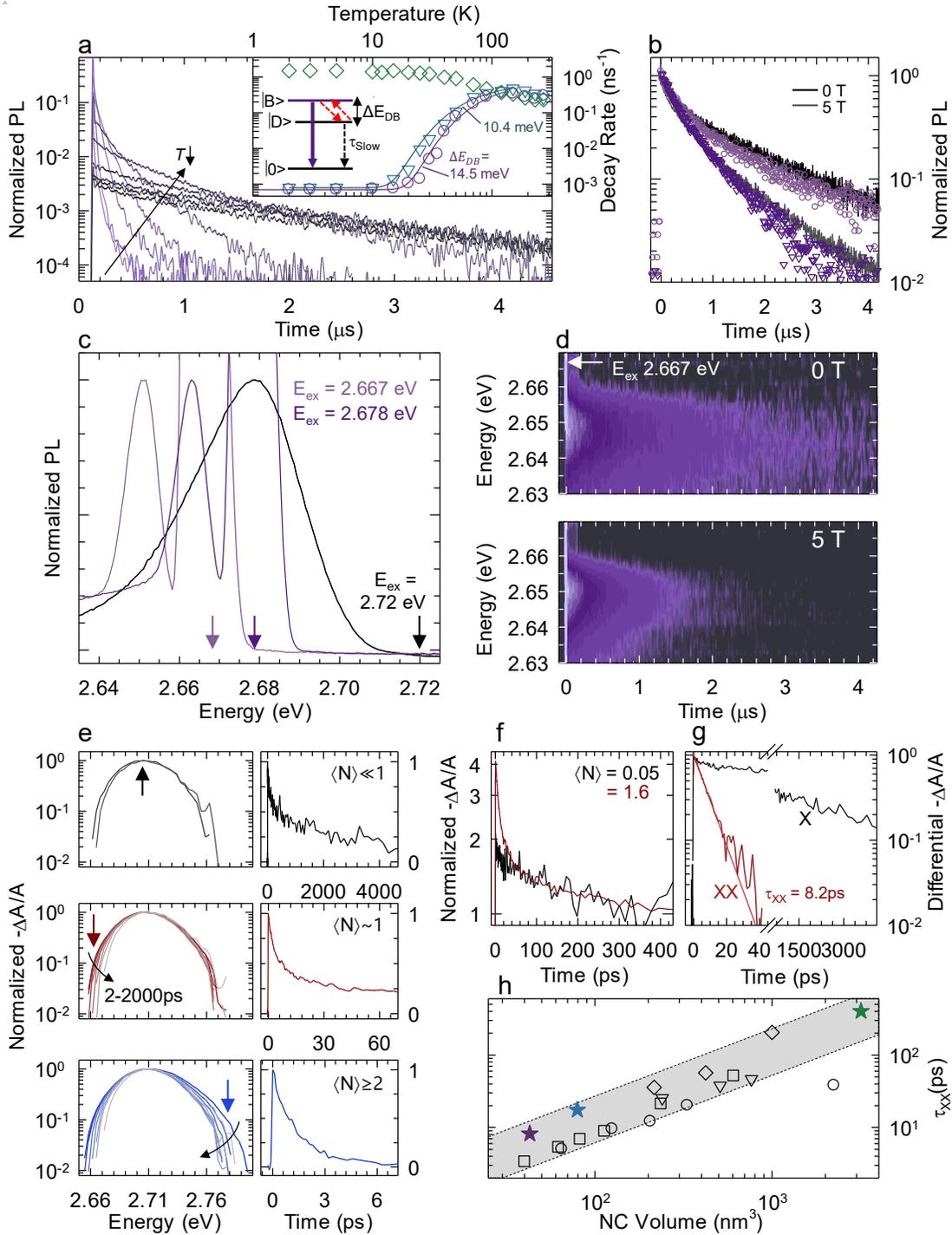

**Figure 4:** (a) Normalized PL decay traces of CsPbBr$_3$ QDs in the temperature range 2 - 300 K for CsPbBr$_3$ QDs with $d$ = 3.5 nm. Inset: the extracted PL decay rates (circles) compared to the decay rates for $d$ = 4.3 nm QDs (triangles) and $d$=14.7 nm nanocrystals (diamonds). The solid lines are the fit with a 3-level model for the two smaller QDs. (b) Normalized PL decays at 5 K in the absence (black line) and in the presence (grey line) of a 5 T magnetic field. The circles and triangles data points represent the PL decay at 0 T and 5 T, respectively, extracted from the FLN contour in 'd'. (c) Normalized PL spectra collected in FLN mode at 5 K for different excitation energies as indicated in figure. The shaded area corresponds to the ensemble PL spectrum excited in non-resonant conditions at 2.72 eV. (d) Time and spectrally resolved FLN decay curves collected at 5 K in the absence (top) and in the presence (bottom) of a 5 T magnetic field. (e) Transient absorption spectra (logarithmic scale) at increasing time (t = 2, 4, 6, 10, 20, 50, 500, 2000 ps) after the excitation pulse for progressively larger average exciton population ⟨N⟩ showing the emergence of bi- and multiexciton spectral contributions. The respective decay traces taken at the energies indicated in the figure are shown in the right-hand panels highlighting gradually faster decay with increasing ⟨N⟩. (f) TA dynamics measured at 2.71 eV and normalized on the single exciton decay tail for ⟨N⟩=0.05 and 1.6. (g) Differential TA curve extracted from panel 'f' representing the biexciton dynamics. The solid line is the fit to single-exponential decay. The single exciton TA trace is also reported in black for direct comparison. h) Measured biexciton lifetime ($\tau_{XX}$) of CsPbBr$_3$ nanocrystals as a function of nanocrystal volume together with other values reported in literature and adapted from ref[62] (squares), ref[66] (circles), ref[59] (triangles), and ref[55] (diamonds). The shaded area is a guide for the eye.

For a further confirmation of the obtained values, we also prepared 4.3 nm and 14.7 nm size CsPbBr$_3$ QDs, using the standard hot-injection synthesis approaches (**Figure S24**). Increasing the QD size to $d$ = 4.3 nm reduced $\Delta_{DB}$ to 10.4 meV. The same analyses, run on the much larger CsPbBr$_3$ nanocrystals with $d$ = 14.7 nm, with size beyond the quantum confinement regime, revealed that the exciton fine structure is inverted with the lower-lying exciton state being a bright triplet, resulting in increasingly faster recombination dynamics with decreasing temperature (diamonds in the inset of **Figure 4a**). Both such findings are in accordance with theory,[47] and with previous reports.

Further independent confirmation of the exciton fine structure of small QDs came from the study of the PL decay in a magnetic field and from fluorescence line narrowing (FLN) measurements. Specifically, as shown in **Figure 4b**, the application of an external magnetic field induced mixing between the dark and bright exciton states, leading to a faster PL decay even at low temperature.[67,68] The $\Delta_{DB}$ energy splitting can also be investigated directly from spectral measurements as the Stokes shift between the so-called zero-phonon line (ZPL), corresponding to the emission from the |D⟩ state without the assistance of phonons, and a ultranarrow excitation line.[69-71] In QD ensembles, where the PL width is broadened by both homogeneous and inhomogeneous effects, respectively due to thermal effects and particle size inhomogeneity, this is performed by operating in FLN configuration at cryogenic temperatures, that is, by exciting the sample at the low energy edge of the PL peak. Such resonant excitation conditions result in size selective excitation only of the large particles subset in the ensemble, thus allowing the intrinsic signatures of exciton fine structure to emerge. This method has been largely used for studying chalcogenide QDs[72] but so far was never applied to halide perovskite QDs. **Figure 4c** shows the FLN spectra at 5 K together with the corresponding ensemble PL when excited at 2.72 eV, slightly above the 1S absorption edge. Notably, whereas the ensemble PL was the typical Gaussian peak due to QD size distribution, pumping on the red tail of the PL resulted in a sharp ZPL peak redshifted from the excitation line by $\Delta_{DB}$=15 meV. In fact, as the QD excitation is due to optical coupling between the ground state and the optically accessible bright exciton, the observed Stokes shift of the ZPL provides direct access to the energy splitting which closely matches with the value obtained from the analysis of PL dynamics. In order to unambiguously ascribe the FLN peak to the ZPL, we further measured the time-resolved FLN spectra reported in **Figure 4d**, showing that the ZPL in fact exhibited the µs-long decay of the dark exciton measured in ensemble mode. When an external magnetic field was applied, the time dynamics of ZPL accelerated identically to the ensemble kinetics (**Figure 4b**). This further confirms our assignment and adds a valuable experimental assessment of the exciton fine structure in the strong confinement regime.

Besides the effects on the fine structure of single excitons, for strongly confined QDs it is important to study the multi-excitonic photophysics, which are typically largely dominated by Auger-type non-radiative processes[72]. To this end, we used fluence-dependent TA spectroscopy to extract the kinetic and spectral components of single-, bi-, and multi-excitonic states in our CsPbBr$_3$ QDs. **Figure 4e** shows the normalized TA bleaching spectra of 3.5 nm sized CsPbBr$_3$ QDs at increasing probe delays in the 2-2000 ps time interval and at progressively higher average exciton population ⟨N⟩, exploring the single-exciton (X), bi-exciton (XX), and multi-exciton (MX) regimes. For ⟨N⟩~0.05, when the bi-exciton population is negligible, the TA spectra exhibited a peak at 2.71 eV due to the filling of band-edge states which decayed in ~2.5 ns, consistent with the PL decay measured at vanishingly low excitation power, together with a ~1.3 ns minor initial contribution likely due to residual electron trapping, consistent with the measured PLQY = 60%. At higher fluences, the biexcitons emerged in the TA spectrum as a lower-energy shoulder - consistent with the attractive nature of the biexciton in CsPbBr$_3$ nanocrystals[66, 73] - which decayed rapidly with a characteristic time of ~9 ps essentially dominated by Auger nonradiative processes. Increasing further the excitation fluences, the TA spectra showed an additional contribution on the high-energy side of the 1S bleach peak. This contribution decayed in approximately 1.4 ps and was attributed to higher order multiexcitons, consistent with previous observations on CsPbBr$_3$ nanocrystals.[59, 74] The detailed analysis of the TA dynamics is presented in **Figure 4f,g** where the biexciton component was extracted according to the procedure introduced by Klimov *et al.*[75] subtracting the contributions of lower order excitons. The measured biexciton decay time was $\tau_{XX}$ = 8.2 ps. Recent reports[55, 62, 66, 76] have shown the general agreement of CsPbBr$_3$ nanocrystals with the universal volume-scaling law of $\tau_{XX}$, similarly to other NC compositions with both direct and indirect energy gaps[77]. To put our data in perspective, in **Figure 4h** we report the biexciton lifetime of the CsPbBr$_3$ QDs obtained from the transformation of (K$_{0.18}$Cs$_{0.82}$)$_4$PbBr$_6$ nanocrystals ($d$=3.5 nm) together with two control samples produced by hot injection ($d$ = 4.3, 14.7 nm) showing good agreement with the general volume scaling.

In summary, we demonstrated a facile approach to synthesize strongly quantum confined cubic-shape CsPbBr$_3$ QDs below 3.5 nm size. The method is based on a slow dissolution/recrystallization reaction of (K$_{0.18}$Cs$_{0.82}$)$_4$PbBr$_6$ nanocrystals upon reacting with PbBr$_2$. The synthesis of (K$_{0.18}$Cs$_{0.82}$)$_4$PbBr$_6$ nanocrystals was performed through a K-exchange reaction on as-synthesized Cs$_4$PbBr$_6$ nanocrystals. The K$^+$ incorporation into Cs$_4$PbBr$_6$ nanocrystals occurs spontaneously at room temperature and stabilizes the crystal lattice, allowing the slowdown of the recrystallization reaction kinetics and thus the CsPbBr$_3$ QDs growth control. By varying the amount of (K$_{0.18}$Cs$_{0.82}$)$_4$PbBr$_6$ nanocrystals in the recrystallization reaction and thus the reaction time we can precisely tune the size and PL emission of the CsPbBr$_3$ QDs in a range from ~ 444 to 462 nm. The as-synthesized CsPbBr$_3$ QDs show good long-term stability, narrow size distribution and excellent optical properties, including tunable absorption and PL emission and with PLQY ~ 60 % (for 3.5 nm size). In-depth spectroscopic analysis provides the first investigation of the exciton fine structure using *cw* and time-resolved line-narrowing techniques in CsPbBr$_3$ QDs and confirms the equivalence of these particles to those synthesized with more traditional methods, further expanding our knowledge on the photophysics of strongly confined lead halide perovskite QDs.

## ASSOCIATED CONTENT

**Supporting Information**. Detailed description of experimental methods and additional data including nanocrystals synthesis, TEM and STEM characterization, XRD and profile fitting, elemental mapping (EDS), elemental analysis (ICP), $^1$H NMR characterization and computational methods.

## AUTHOR INFORMATION

**Corresponding Author**

*Liberato Manna - Nanochemistry, Istituto Italiano di Tecnologia, 16163 Genova, Italy; orcid.org/0000-0003-4386-7985; Email: liberato.manna@iit.it

*Sergio Brovelli - Dipartimento di Scienza dei Materiali, Università degli Studi di Milano-Bicocca, 20125 Milano, Italy; orcid.org/0000-0002-5993-855X; Email: sergio.brovelli@unimib.it

**Notes**

The authors declare no competing financial interest.

## ACKNOWLEDGMENT

C.O.M. acknowledges the Xunta de Galicia for a PhD scholarship (ED481A-2021/318). M.L.Z., A.F. and S.B. acknowledge the financial support of the Horizon Europe EIC Pathfinder program, project 101098649 - UNICORN and of the Italian Ministry of Research PRIN program, project IRONSIDE. The authors wish to thank Alexander L. Efros for valuable suggestions. L.P. acknowledges the support from the Spanish Ministerio de Ciencia e Innovación through Ramón y Cajal grant (RYC2018-026103-I) and the Spanish State Research Agency (Grant No. PID2020-117371RA-I00), the grant from the Xunta de Galicia (ED431F2021/05). G. S. and L. M. acknowledge funding from the Project IEMAP (Italian Energy Materials Acceleration Platform) within the Italian Research Program ENEA-MASE (Ministero dell'Ambiente e della Sicurezza Energetica) 2021-2024 "Mission Innovation" (agreement 21A033302 GU n. 133/5-6-2021).

# Supporting information

# Ultrasmall CsPbBr$_3$ Blue Emissive Perovskite Quantum Dots using K-alloyed Cs$_4$PbBr$_6$ Nanocrystals as Precursors

*Clara Otero Martínez[1,2,‡], Matteo L. Zaffalon[3,‡], Yurii Ivanov[4], Nikolaos Livakas[2,5], Luca Goldoni[6], Giorgio Divitini[4], Sankalpa Bora[7], Gabriele Saleh[2], Francesco Meinardi[3], Andrea Fratelli[3], Sudip Chakraborty[7], Lakshminarayana Polavarapu[1], Sergio Brovelli[3]\*, Liberato Manna[2]\**

[1] *CINBIO, Department of Physical Chemistry, Materials Chemistry and Physics Group, Universidade de Vigo, Campus Universitario As Lagoas-Marcosende, 36310 Vigo, Spain*

[2] *Nanochemistry, Istituto Italiano di Tecnología, Via Morego 30, 16163 Genova, Italy*

[3] *Dipartimento di Scienza dei Materiali, Università degli Studi di Milano-Bicocca, Via R. Cozzi 55, 20125, Milano, Italy*

[4] *Electron Microscopy and Nanoscopy, Istituto Italiano di Tecnología, Via Morego 30, 16163, Genova, Italy*

[5] *Dipartimento di Chimica e Chimica Industriale, Università di Genova, 16146 Genova, Italy*

[6] *Material Characterization Facility, Istituto Italiano di Tecnologia Via Morego 30, 16163 Genova, Italy*

[7] *Harish-Chandra Research Institute (HRI) Allahabad, Jhunsi, Prayagraj, 211019, India*

‡ Equal Contribution

AUTHOR INFORMATION

**Corresponding Author**

*Liberato Manna - Nanochemistry, Istituto Italiano di Tecnologia, 16163 Genova, Italy; orcid.org/0000-0003-4386-7985; Email: liberato.manna@iit.it

*Sergio Brovelli - Dipartimento di Scienza dei Materiali, Università degli Studi di Milano-Bicocca, 20125 Milano, Italy; orcid.org/0000-0002-5993-855X; Email: sergio.brovelli@unimib.it

## Materials and methods

**Materials:** Cesium carbonate ($Cs_2CO_3$, 99,9%), potassium acetate ($CH_3COOK$, > 98 %), lead bromide ($PbBr_2$, >98%), lead acetate trihydrate ($Pb(CH_3COO)_2 \cdot 3H_2O$, 99.99%), benzoyl bromide ($C_6H_5COBr$, 97%), octadecene ($C_{18}H_{36}$, 90%) oleic acid ($C_{18}H_{34}O_2$, 90%), oleylamine ($C_{18}H_{37}N$, 70%), toluene ($C_7H_8$, >99.8%), acetonitrile ($CH_3CN$, > 99.9 %), ethyl acetate ($C_4H_8O_2$), deuterated dimethyl sulfoxide-$d_6$ ($(CD_3)_2SO$,) and deuterated toluene-$d_8$ ($C_6D_5CD_3$, 99.95%) were purchased from Merck. All chemicals were used without further purification.

*Precursor solutions preparation*

**$PbBr_2$ precursor solution 1 ($Cs_4PbBr_6$ synthesis).** $PbBr_2$ (60 mg), 5 mL of octadecene, 1.5 mL of oleylamine and 200 μL of oleic acid were loaded in a 40 mL vial. The solution was heated-up to 120 °C under stirring for 30 min. After this time, the salt was completely dissolved and the solution was cooled down.

**Cs-OL solution ($Cs_4PbBr_6$ synthesis).** $Cs_2CO_3$ (400 mg) and 8 mL of oleic acid were loaded in a 25 mL vial. The solution was heated up to 100 °C under stirring and degassed under vacuum for 1 hour. After this time, the salt was completely dissolved solution was cooled down.

**K-OL solution (cation exchange reaction).** First, potassium acetate (250 mg) and 10 mL of oleic acid were loaded in a 25 mL vial to prepare a 0.25 M K-OL solution. The mixture was heated-up to 100 °C under stirring and degassed under vacuum during 1 hour. After that, the salt was completely dissolved and the solution was cooled down. For the K-exchange reactions, 1 mL of the 0.25 M K-OL solution was diluted in 9 mL of toluene to prepare a ≈ 25 mM K-OL solution.

**PbBr₂ precursor solution 2 (CsPbBr₃ QDots synthesis).** PbBr₂ (367 mg), 5 mL of toluene, 2.5 mL of oleylamine and 2.5 mL of oleic acid were loaded in a 25 mL vial. The solution was heated up to 120 °C under stirring for 1 hour. After this time, the salt was completely dissolved and the solution was cooled down.

**Pb-Cs-OL stock solution (large CsPbBr₃ nanocrystals synthesis).** Lead acetate trihydrate (760 mg), Cs₂CO₃ (160 mg) and oleic acid (15 mL) were loaded in a 25 mL vial. The solution was heated up to 100 °C under stirring and degassed for 1 hour. After this time, the precursors salts were completely dissolved and the solution was cooled down.

**DDA stock solution (large CsPbBr₃ nanocrystals synthesis).** Didodecylamine (4.43 g) and toluene (10 mL) were loaded in a 25 mL vial. To dissolve the salt, the mixture was sonicated for 10 min at 50 °C in a sonication bath.

*Nanocrystal synthesis*

**Cs₄PbBr₆ nanocrystals synthesis.** The synthesis of Cs₄PbBr₆ nanocrystals was performed accordingly a previously reported method[1] with slight modifications. Briefly, the **PbBr₂ precursor solution 1** previously prepared was heated up to 80 °C in a 20 mL vial under stirring and N₂ atmosphere. When the solution reached the desired temperature, 0.75 mL of the Cs-OL solution was swiftly injected. After ≈ 30 s, the solution became white-turbid indicating the formation of the Cs₄PbBr₆ nanocrystals and the reaction was quenched in an ice-bath. The solution was purified by centrifugation (6000 rpm, 5 min) without use of antisolvents and the precipitated was collected and redispersed in 1 mL of toluene.

**(K₀.₁₈Cs₀.₈₂)₄PbBr₆ nanocrystals synthesis.** The synthesis of (K₀.₁₈Cs₀.₈₂)₄PbBr₆ nanocrystals was performed by 3 cycles of cation exchange reaction in air conditions and room temperature. In each cycle, 1 mL of the 25 mM K-OL solution in toluene was added to the as-synthesized Cs₄PbBr₆ nanocrystals solution in toluene in a 7 mL vial. Then, 5 mL of ethyl acetate were added to the colloidal solution to induce the precipitation of the nanocrystals. The solution was centrifuged (6000 rpm, 5 min) and the precipitate was collected and redispersed in a 1 mL oleylamine & oleic acid ligand solution in toluene of 2.5 mM concentration to recover the losses of ligands during the washing with antisolvent. This process was repeated 2 more times to complete the 3 cycles of K-exchange.

**CsPbBr₃ QDs synthesis.** The synthesis of CsPbBr₃ QDs was performed by a dissolution-recrystallization reaction of (K₀.₁₈Cs₀.₈₂)₄PbBr₆ nanocrystals when reacted with PbBr₂. Briefly, 5 mL of toluene and the desired volume of $(K_{0.18}Cs_{0.82})_4PbBr_6$ nanocrystals were loaded in a 7 ml vial. The solution was heated up to 100 °C under stirring and $N_2$ atmosphere. When the solution reached the desired temperature, 1 mL of the **PbBr₂ precursor solution 2** was swiftly injected. After the corresponding reaction time, the solution was quenched in an ice-water bath. The specific volumes of $(K_{0.18}Cs_{0.82})_4PbBr_6$ nanocrystals and the corresponding reaction time are described in **Table S1**.

CsPbBr₃ QDots were purified by adding subsequently 0.5 mL of EtOAc and 0.5 mL of acetonitrile to 1 mL of crude solution. After the addition of the solvents the crude solution became turbid, meaning that the QDots lost the colloidal stability and the solution was quickly centrifuged (4000 rpm, 2 min). The supernatant was discarded and the precipitate was redispersed in 1 mL of toluene.

**Table S1.** Reaction times and $(K_{0.18}Cs_{0.82})_4PbBr_6$ nanocrystals volumes employed to obtained CsPbBr₃ QDs with their corresponding excitonic peak.

| Reaction time | Volume $(K_{0.18}Cs_{0.82})_4PbBr_6$ nanocrystals (mL) | Excitonic peak |
|---|---|---|
| 5 | 0.05 | 420.0 |
| 10 | 0.1 | 425.7 |
| 15 | 0.15 | 430.7 |
| 30 | 0.25 | 436.2 |
| 50 | 0.35 | 439.7 |
| 100 | 0.45 | 442.3 |
| 135 | 0.5 | 448.0 |
| 160 | 0.55 | 452.5 |

**4.3 and 15 nm CsPbBr₃ nanocrystals synthesis**. The synthesis of CsPbBr₃ nanocrystals was performed according previously reported methods with slight modifications.[2] Briefly, octadecene (9 mL), **Pb-Cs-OL sock solution** (1.5 mL) and **DDA stock solution** (1.5 mL) are loaded in a 25 mL vial and heated up under $N_2$ and stirring to 50 and 160 °C to obtain 4.3 and 15 nm size respectively. Then, a solution of benzoyl bromide (50 μL) in toluene (500 μL) was swiftly injected.

After 1 min, the reaction was quenched in an ice-water bath. When the reaction was cooled down, 200 μL of oleylamine was added to the crude solution to obtain the OLAm & OL surface capping composition. The colour change to a bright green indicates the oleylammonium binding to the surface. The solutions were then purified by centrifugation (6000 rpm, 5 min) after adding 6 mL of EtOAc to 3 mL of the crude solution. Finally, the precipitate obtained was redispersed in 5 mL of decane for the spectroscopy measurements.

*NMR characterization*

NMR experiment were acquired at 298 K, on a Bruker Avance III 600 MHz (600.13 MHz) spectrometer, fit with a 5 mm QCI cryoprobe. Prior to the acquisition, matching and tuning and, line shape resolution, were automatically adjusted. The 90° pulse calculated on each tube by using Bruker's automatic routines.[3] $^1$H NMR spectra in toluene-$d_8$ were performed without steady scan, with 256 scans and 65536 points of digitalization, an inter pulses delay of 30 s, at a fixed receiver gain (18), on a spectral width of 20.83 ppm centered at 6.18 ppm.

For the *$^1$H quantitative* NMR spectra in DMSO-$d_6$ used for concentration measurement with PULCON method, identical acquisition parameters were employed except for the number of transients (64).

Prior to the Fourier transform an exponential smoothing function equivalent to 0.3Hz was applied to FIDs. Spectra were manually phased and automatically baseline corrected.

**Sample preparation.** For the NMR characterization, $Cs_4PbBr_6$ and $(K_{0.18}Cs_{0.82})_4PbBr_6$ nanocrystals solutions in toluene (1 mL) were purified with EtOAc (6 mL) by centrifugation (6000 rpm, 5 min). Then, the precipitates were dried with a $N_2$ flow and redispersed in deuterated toluene (500 μL). The colloidal solution was loaded into a 5mm disposable sample Jet tubes (Bruker). For the ligand quantification, the colloidal solutions already characterized in deuterated toluene, where dried using a $N_2$ flow and the resulting material was dissolved in deuterated DMSO (200 μL) and loaded into a 3 mm disposable sampleJet tube.

**Ligand quantification.** The concentration of both ligands ([OLAm & OL]) was quantified using *quantitative* NMR and PULCON (PUlse Length-based CONcentration determination) method[4], comparing the integrated peak intensity of ligand signals to that of standard external solution of dimethylsulfone (10 mM, TraceCERT®) in DMSO-$d_6$, after normalizing each signal to the number of 1H resonances generating the peak. The peaks used for quantification are the pseudo triplet at 2.75 ppm for OLAm and triplet at 2.1 ppm for OL.

Total ligand concentration = [OLAm + OL] = [OLAm]$_{\text{measured by PULCON}}$ + [OL] $_{\text{measured by PULCON}}$      *SEq. 1*

**Inductively Coupled Plasma−Optical Emission Spectroscopy (ICP-OES).**

After the ligand quantification, the Pb concentration of the nanocrystals dissolved in deuterated DMSO-$d_6$ was obtained by ICP-OES on an aiCAP 6000 spectrometer (Thermo Scientific). Prior the measurement, 50 µL of the DMSO – $d_6$ sample were diluted to 10 mL in an aqueous solution of aqua regia (1:10) and subjected to an acid digestion overnight.

**Ligand density**

The ligand density of the samples was obtained by calculating the ratio of the ligand concentration and nanocrystal surface in the DMSO-$d_6$ solution:

$$Ligand\ density = \frac{[Ligand](mL^{-1})_{PULCON}}{Total\ NC\ surface\ (nm^2/mL)} \qquad SEq.\ 2$$

The total nanocrystal surface was determined through the Pb concentration in the DMSO-$d_6$ sample (measured by ICP) and the size distribution of the nanocrystals considering a unit cell volume of 2.82 and 2.78 nm$^3$ for Cs$_4$PbBr$_6$ and (K$_{0.18}$Cs$_{0.82}$)$_4$PbBr$_6$ nanocrystals respectively.

*Computational Methodology*

In order to substitute K-atoms in the Cs$_4$PbBr$_6$ lattice we have incorporated the Site-Occupation Disorder (SOD) program.[5] Subsequently, we have systematically determined various electronic properties, including the density of states, band structure, and optical spectra for the studied configurations. We have conducted comprehensive electronic structure calculations based on first principles using the density functional theory (DFT) framework [5, 6] within the Vienna ab-initio simulation package (VASP).[7] Throughout the geometry optimization and electronic structure calculations, we have applied the projected-augmented-wave (PAW) formalism[8] and employed the generalized gradient approximation (GGA) for the exchange-correlation functional.[9] For the bulk geometry optimization, we utilized a plane wave basis set with a 400 eV energy cutoff and a 4x4x3 Monkhorst Pack k-points sampling scheme.[10] We ensured that all considered systems were fully relaxed according to the minimum-energy criteria, continuing the optimization until the Hellman-Feynman force dropped below 0.02 eV/Å. For density of states calculation, we took a denser grid k-points using 9x9x6 Monkhorst pack scheme and specified the number of energy grid points to 3000 while keeping the other parameters same. The band structure and optical spectra calculations were done using the same parameters but with 4x4x3 k-points.

*Optical characterization*

UV-Vis absorption spectra were carried out using a Varian Cary 300 UV−Vis absorption spectrophotometer (Agilent). The spectra were collected by diluting 50 µL of the sample in toluene in 2.5 mL of hexane in order to collect the spectra until ≈ 250 nm and analyse the features of $Cs_4PbBr_6$ and $(K_{0.18}Cs_{0.82})_4PbBr_6$ nanocrystals. Photoluminescence spectra were obtained on a Varian Cary Eclipse Spectrophotometer (Agilent) using $\lambda_{ex}$= 350 nm ($CsPbBr_3$ QDs) and 270 nm ($Cs_4PbBr_6$ and $(K_{0.18}Cs_{0.82})_4PbBr_6$ nanocrystals). Time resolved photoluminescence spectra at room temperature and PL quantum yield measurements were obtained using an Edinburgh FLS900 fluorescence spectrophotometer. PL decay traces were measured with a pulsed laser diode ($\lambda_{ex}$= 375, pulse width = 50 ps). Quantum yield measurements were acquired using a calibrated integrating sphere with $\lambda_{ex}$ = 350 nm for all of the measurements. All solutions were diluted to an optical density of 0.1 - 0.2 at the excitation wavelength in order to minimize the reabsorption of the fluorophore. Quartz cuvettes with an optical path length of 1 cm were used for all optical analyses.

The in-situ absorption spectra were collected directly from the reaction medium via a transmission dip probe (Anglia Instruments, fiber optic immersion probe AIFDP-12UV200600-2-SS-VHT). The signal was collected with an AvaSpec-2048 spectrophotometer controlled through AvaSoft 8 software, version 8.14. The spectra resolution is 0.5 s per spectrum, integration time 2 ms, 250 averages per spectrum, and smoothing set to 6 pixels. The light source used for the ABS measurements was an AvaLight-DH-S-BAL, while the light.

Temperature-dependent and time-resolved PL measurements were performed on QD thin films drop-cast on quartz substrates and mounted in a closed-circuit He cryostat with optical access and equipped with superconducting coils for magnetic field generation. The excitation source was a pulsed laser at 3.06 eV (405 nm, ~70 ps pulses) and the emitted light was dispersed with a spectrometer and detected with a charge-coupled device for *cw* measurements and with a photomultiplier tube coupled with a time-correlated single-photon counting unit for time-resolved PL measurements (time resolution ~400 ps).

For FLN measurements, a spectrally narrowed (~0.5 nm full width at half maximum) fs-pulsed laser was generated by coupling the output of a tunable optical parametric amplifier to a 1/3 m double-grating Gemini monochromator. The emitted PL was collected with a Horiba Scientific Triax 180 1/2 m spectrograph and detected with a cooled charge-coupled device.

Ultrafast transient absorption spectroscopy measurements were performed on a Helios TA spectrometer from Ultrafast Systems. The laser source was a 10 W Hyperion amplified laser operating at 1.875 kHz and producing ~260 fs pulses at 1030 nm, coupled to an independently tunable APOLLO-Y optical parametric amplifier from the same supplier that produced the excitation pulses at 3.1 eV. After passing the pump beam through a synchronous chopper phase-locked to the pulse train (0.937 kHz, blocking every second pump pulse), the pump fluence on the sample was modulated from 13 $\mu$J cm$^{-2}$ to 1700 $\mu$J cm$^{-2}$. The probe beam was a white light supercontinuum.

*Elemental analysis*

Scanning electron microscopy (SEM) was performed on a HRSEM JEOL JSM-7500LA microscope with a cold field-emission gun (FEG), operating at acceleration voltage of 15 kV. The elemental composition of the samples was obtained by energy-dispersive spectroscopy (EDX, Oxford instrument, X-Max, 80 mm2) operating at 8 mm working distance, acceleration voltage of 15 kV, and 15 sweep count.

Transmission electron microscopy (TEM) images were obtained with a JEOL JEM 1011 transmission electron microscope operating at an acceleration voltage of 100 kV. High-resolution scanning transmission electron microscopy (HR STEM) images were acquired on a probe-corrected ThermoFisher Spectra 30-300 STEM operated at 300 kV. Images were acquired on a High-Angle annular Dark Field (HAADF) detector with a very small beam current of around 10 pA to avoid a beam damages of a few nanometres lead halide perovskite nanocrystals. Convergence angle was set to 25 mrad, it corresponds to sub angstrom electron beam. Compositional maps were acquired using Velox, with a probe current of ~150 pA and rapid rastered scanning. The Energy-Dispersive X-Ray (EDX) signal was acquired on a Dual-X system for a total acquisition angle of 1.76 Sr.

*X-ray Powder Diffraction and profile fitting.*

X-ray powder diffraction measurements were performed on a PANanalytical Empyrean X-ray diffractometer, equipped with a 1.8 kW Cu K$\alpha$ ceramic anode and a PIXcel3D 2 × 2 area detector, operating at 45 kV and 40 mA. Nanocrystals' dispersions were mixed with fumed silica and dried to minimize the preferential orientation. Finally, the powder was placed on a zero-diffraction silicon substrate to perform the measurements. X-ray powder diffraction patterns were refined through Rietveld method employing Fullprof Suite software. The analysis included background extraction followed by a systematic refinement of various parameters involving scale factor, unit cell, instrumental and profile parameters (taking into account shape and asymmetry). Likewise, factors such as atomic coordinates, occupancy and thermal expansion were refined.

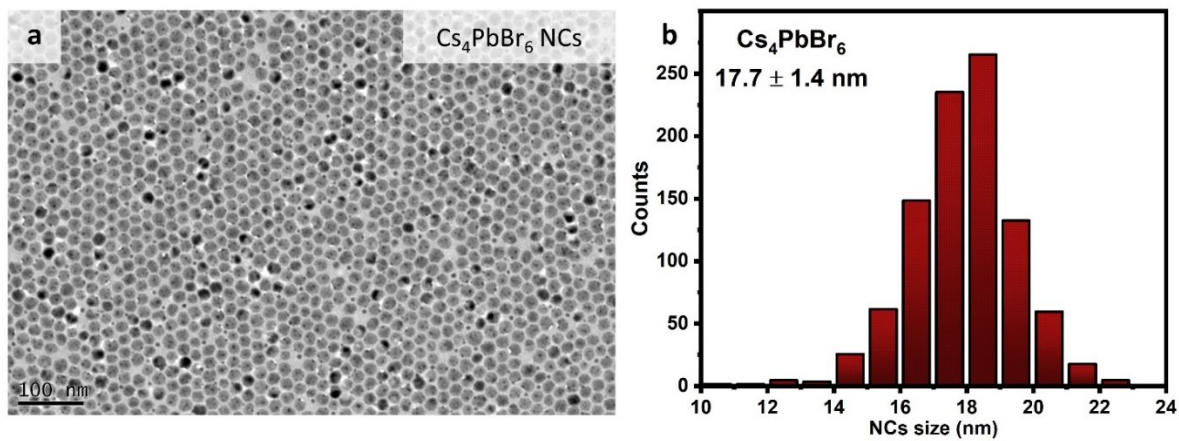

**Figure S1.** (a) TEM images of $Cs_4PbBr_6$ nanocrystals and (b) histogram obtained after the analysis of the corresponding TEM image. The estimated average size was $17.7 \pm 1.4$ nm.

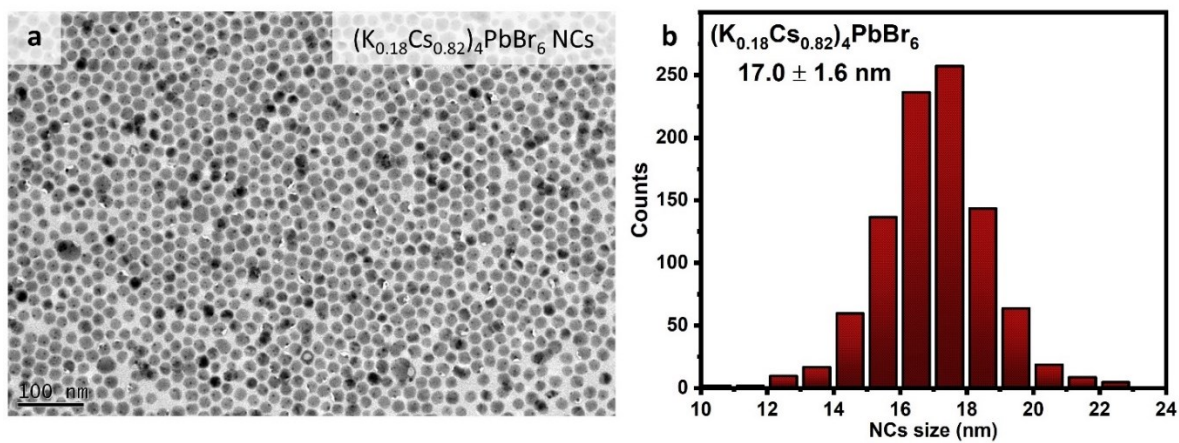

**Figure S2.** (a) TEM images of $(K_xCs_{1-x})_4PbBr_6$ nanocrystals obtained by cation-exchange reaction and (b) histogram obtained after the analysis of the corresponding TEM image. The estimated average size was $17.0 \pm 1.6$ nm.

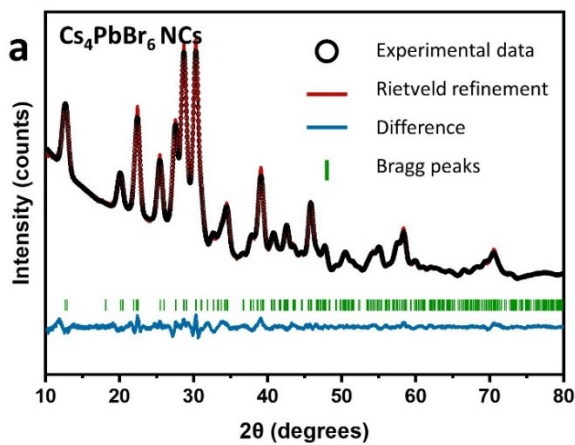

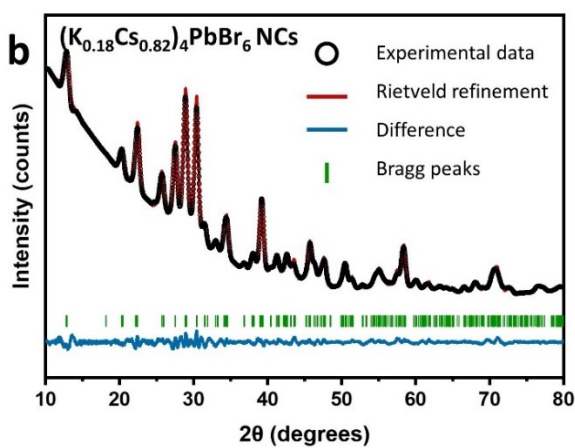

**Figure S3.** Structural characterization. Rietveld refinement of XRD diffraction patterns of $Cs_4PbBr_6$ nanocrystals (a) and K-doped $(K_{0.18}Cs_{0.82})_4PbBr_6$ nanocrystals (b). The black circles and red lines are the experimental and the calculated patterns. The lower line (blue) represents the residuals of the fit, while the vertical bars (green) correspond to the calculated positions of Bragg peaks. Refined parameters of the two structural models are presented in the corresponding tables.

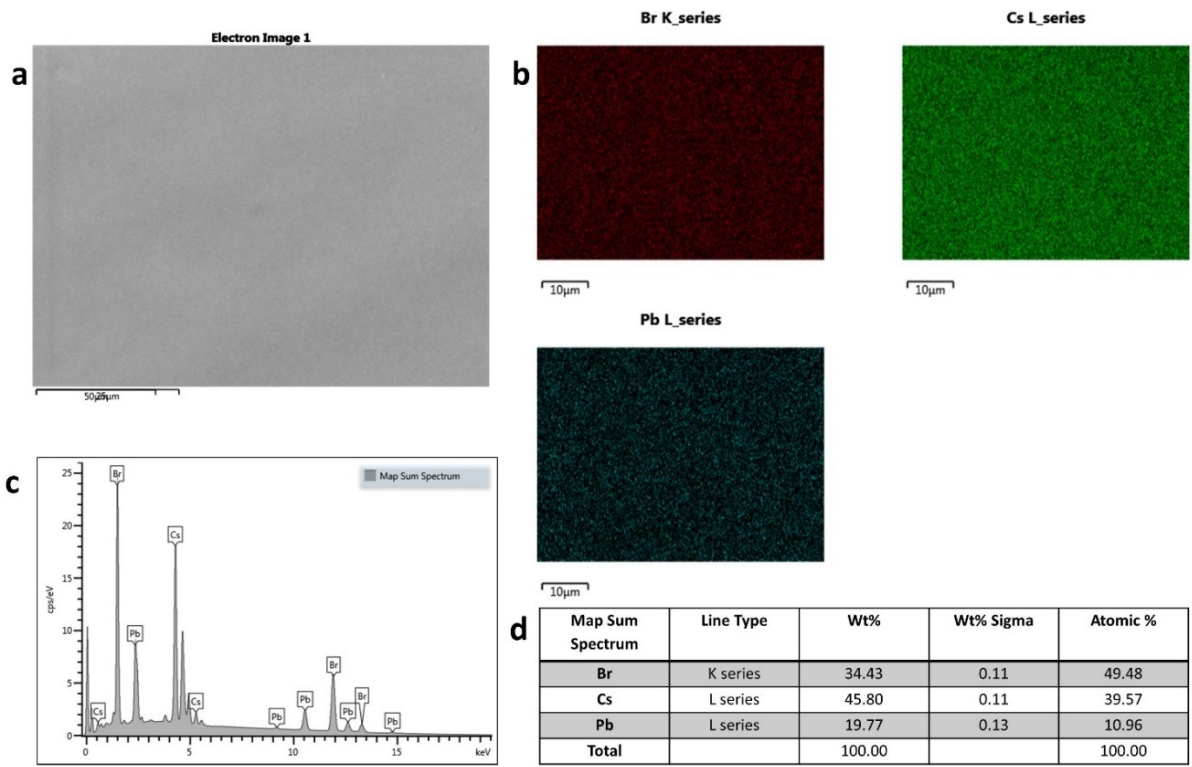

**Figure S4.** a) SEM image of a Cs$_4$PbBr$_6$ nanocrystals film. b) EDS elemental images of Br, Cs and Pb. (c) EDS spectrum of the Cs$_4$PbBr$_6$ nanocrystals film from (a). (d) Summary of the elemental analysis obtained from the EDS spectrum expressed in atomic percentage (%). The scale bar is 50 μm in all the cases.

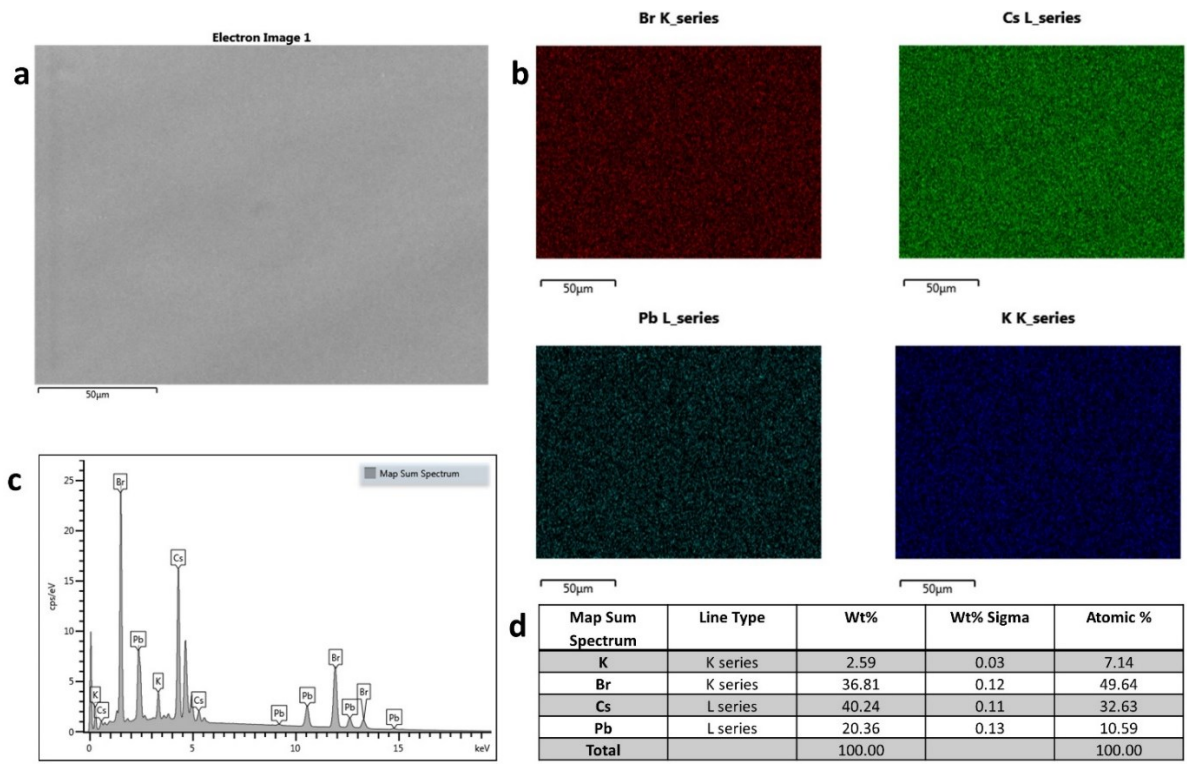

**Figure S5.** a) SEM image of a $(K_xCs_{1-x})_4PbBr_6$ nanocrystals film. b) EDS elemental images of Br, Cs, Pb and K. (c) EDS spectrum of the $(K_xCs_{1-x})_4PbBr_6$ nanocrystals film from (a). (d) Summary of the elemental analysis obtained from the EDS spectrum expressed in atomic percentage (%). The scale bar is 50 μm in all the cases.

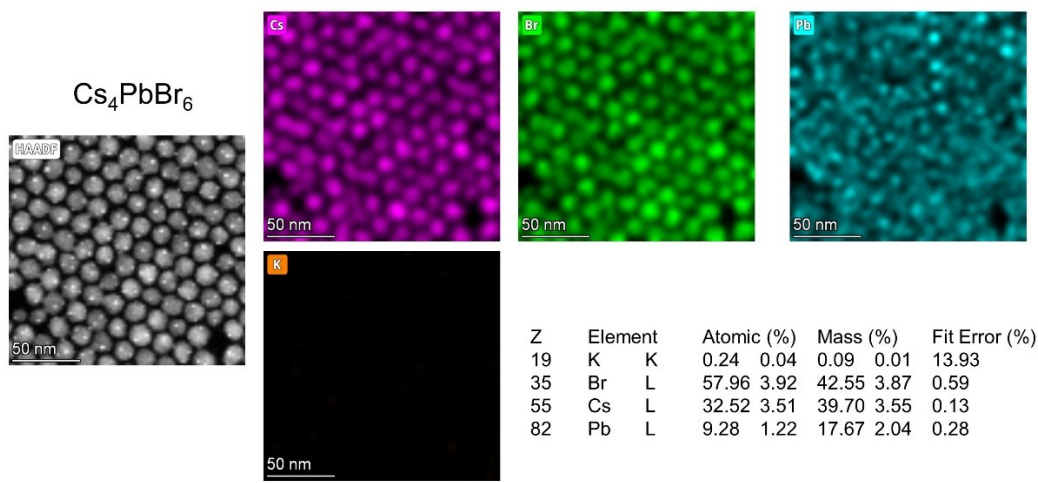

**Figure S6.** a) HAADF-STEM image of Cs$_4$PbBr$_6$ nanocrystals and corresponding EDS elemental images of Cs, Br, Pb and K.

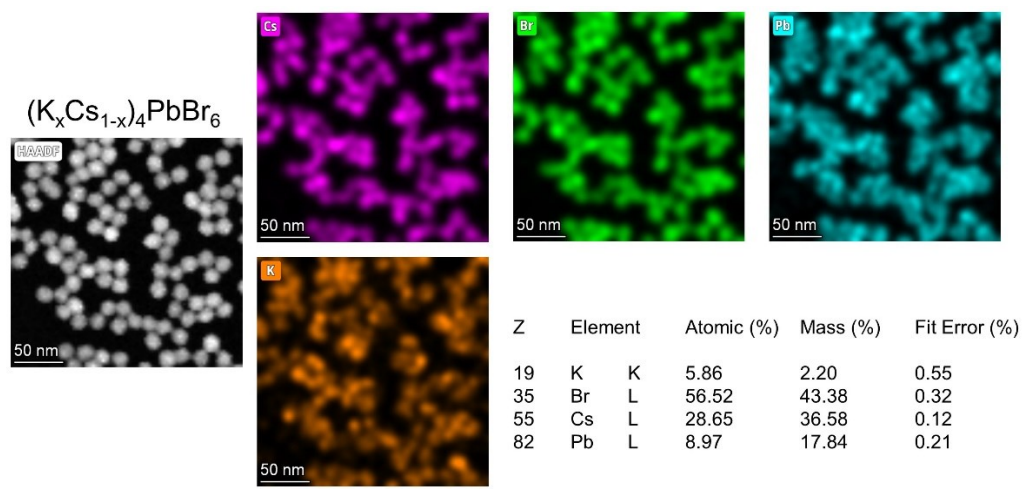

**Figure S7.** a) HAADF-STEM image of (K$_x$Cs$_{1-x}$)$_4$PbBr$_6$ nanocrystals and corresponding EDS elemental images of Cs, Br, Pb and K.

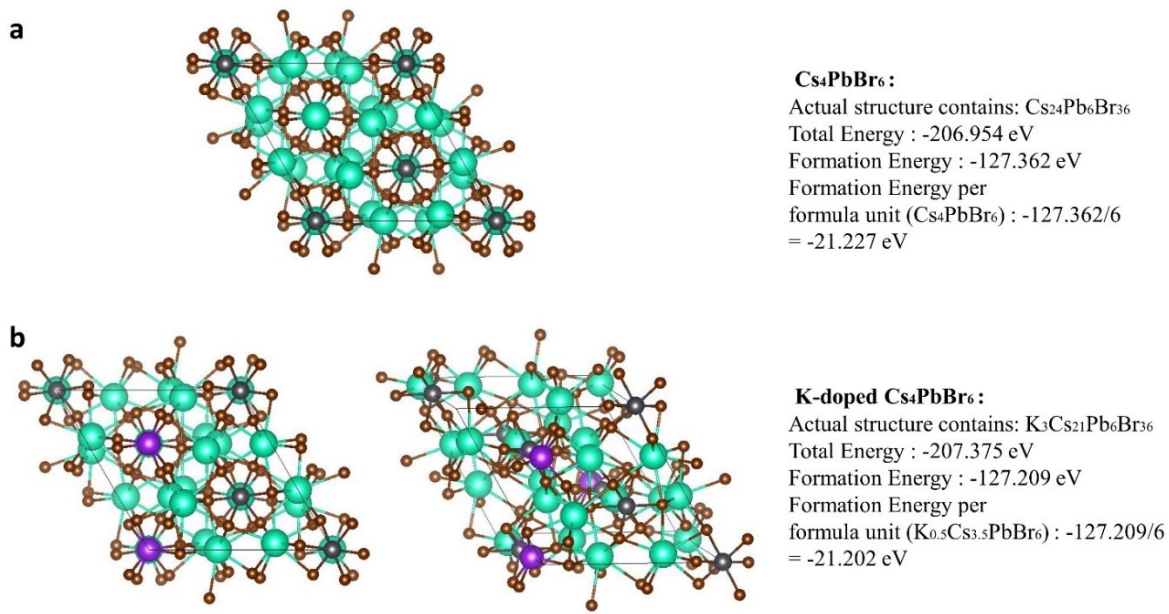

**Figure S8.** Crystal structures and corresponding energies of $Cs_4PbBr_6$ (a) and $(K_{0.125}Cs_{0.875})_4PbBr_6$ (b).

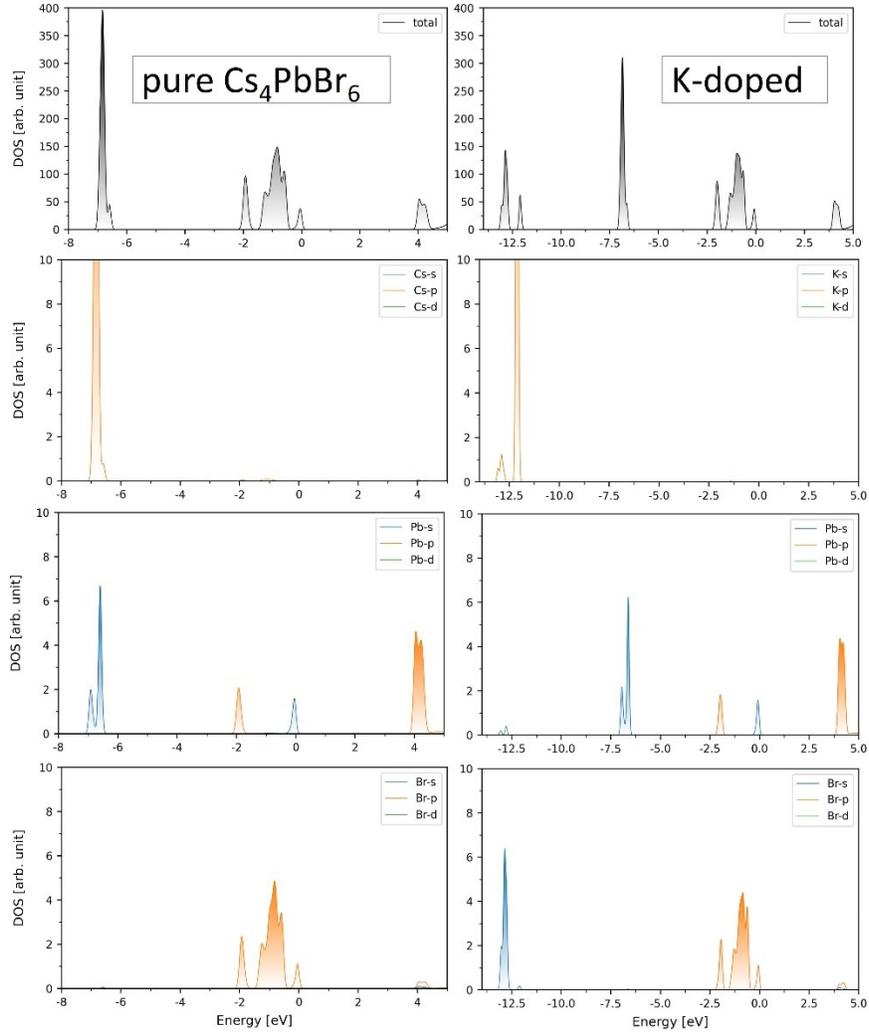

**Figure S9.** Density of states (DOS) for pure (left) and K-doped (right) $Cs_4PbBr_6$ (unit cell formulas: $Cs_{24}Pb_6Br_{36}$ and $K_3Cs_{21}Pb_6Br_{36}$, respectively). The top panels show the total DOS, while those below the atomic orbital contributions of each atom type (Cs for K-doped not shown for sake of space). For projected DOS, each panel shows 2 atoms (2 is the maximum number of symmetry-independent atoms among all atom types in the undoped compound), so that the DOS magnitudes are directly comparable.

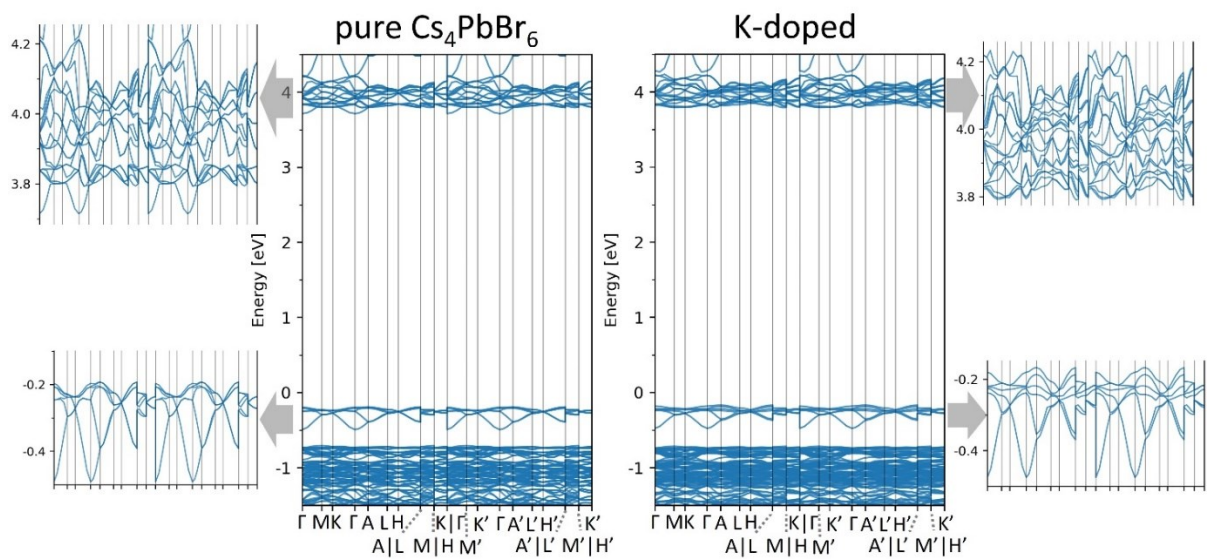

**Figure S10.** Band structure of pure and K-doped $Cs_4PbBr_6$. Enlargements of the valence and conduction bands are shows beside each plot, showing the indirect character of the band gap. Note that for both plots we adopted the reciprocal space path of the K-doped compound, to allow a direct comparison. The latter, indeed, is less symmetric (space group P321 *vs* R-3c of the undoped compound) and thus requires a more extended reciprocal space path.

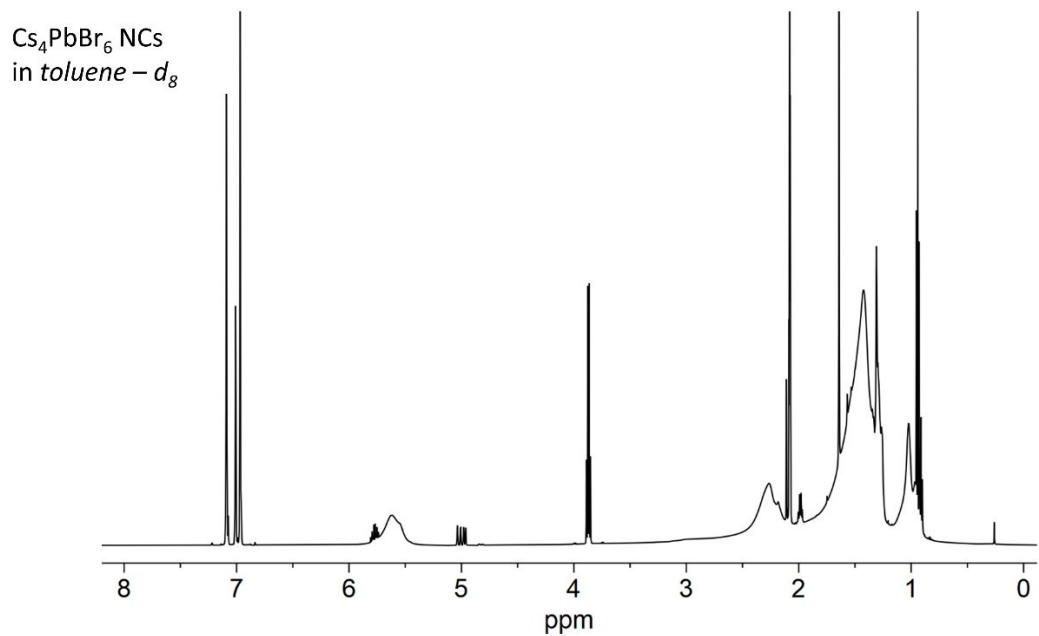

**Figure S11.** $^1$H NMR spectrum of $Cs_4PbBr_6$ in toluene-$d_8$.

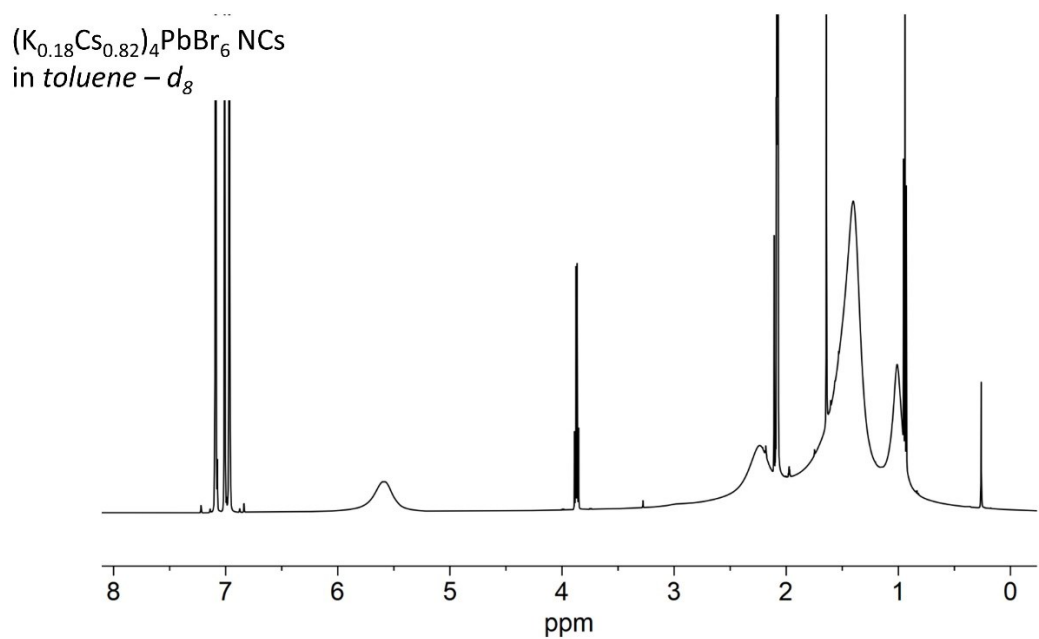

**Figure S12.** $^1$H NMR spectrum of $(K_{0.18}Cs_{0.82})_4PbBr_6$ in toluene-$d_8$.

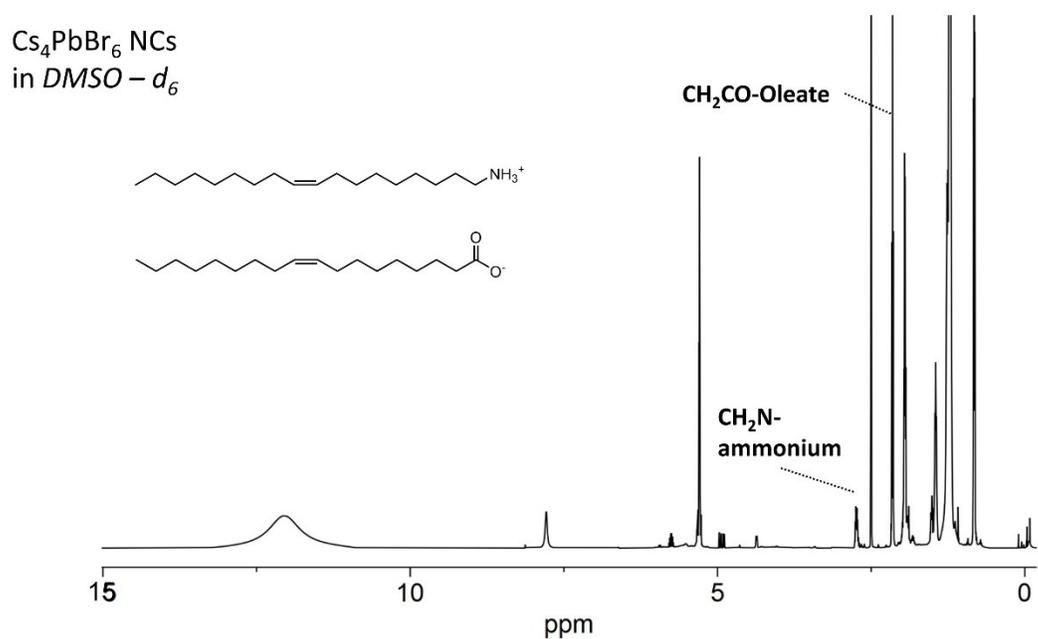

**Figure S13.** $^1$H NMR quantitative spectrum of Cs$_4$PbBr$_6$ in DMSO-d$_6$ + 5 μL (2.5% v/v) of TFA. Peaks used for quantification are shown.

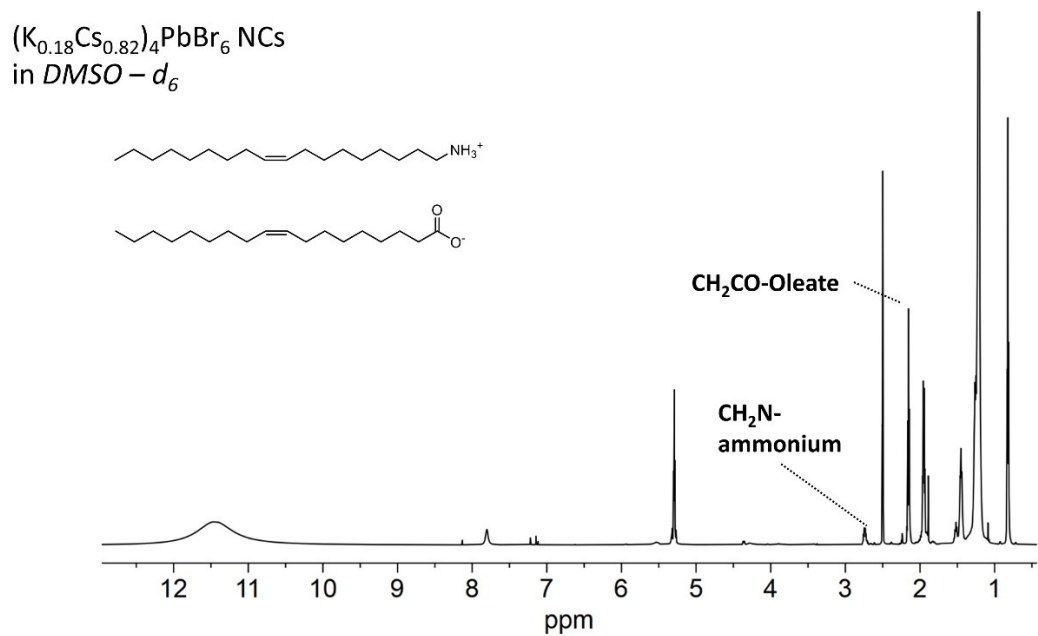

**Figure S14.** $^1$H NMR quantitative spectrum of (K$_{0.18}$Cs$_{0.82}$)$_4$PbBr$_6$ in DMSO-d$_6$ + 5 μL (2.5% v/v) of trifluoro acetic acid (TFA). Peaks used for quantification are shown.

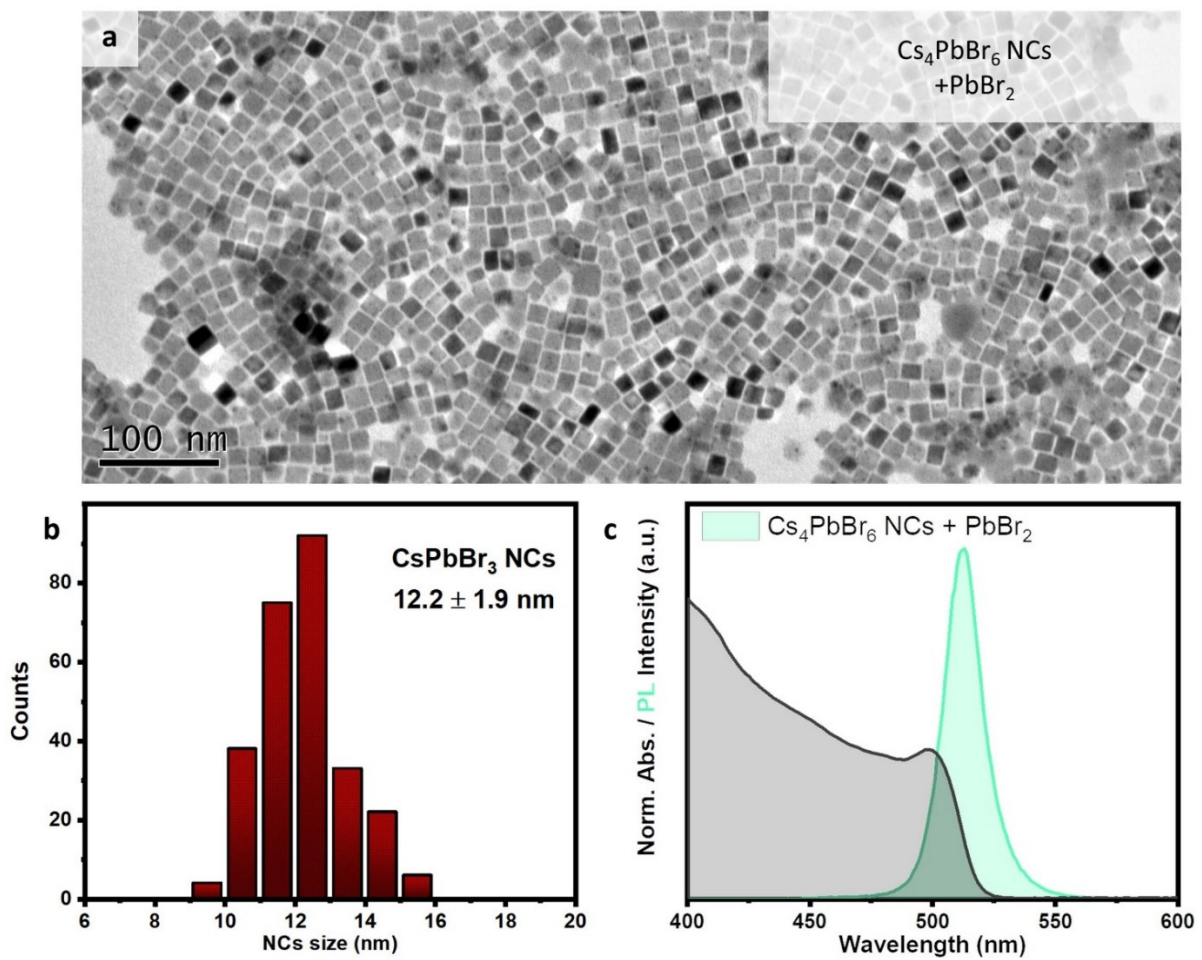

**Figure S15.** (a) TEM image of CsPbBr$_3$ nanocrystals obtained by the reaction of Cs$_4$PbBr$_6$ nanocrystals with PbBr$_2$. (b) Histogram obtained after the analysis of the corresponding TEM image. The estimated average size was 12.2 ± 1.9 nm.

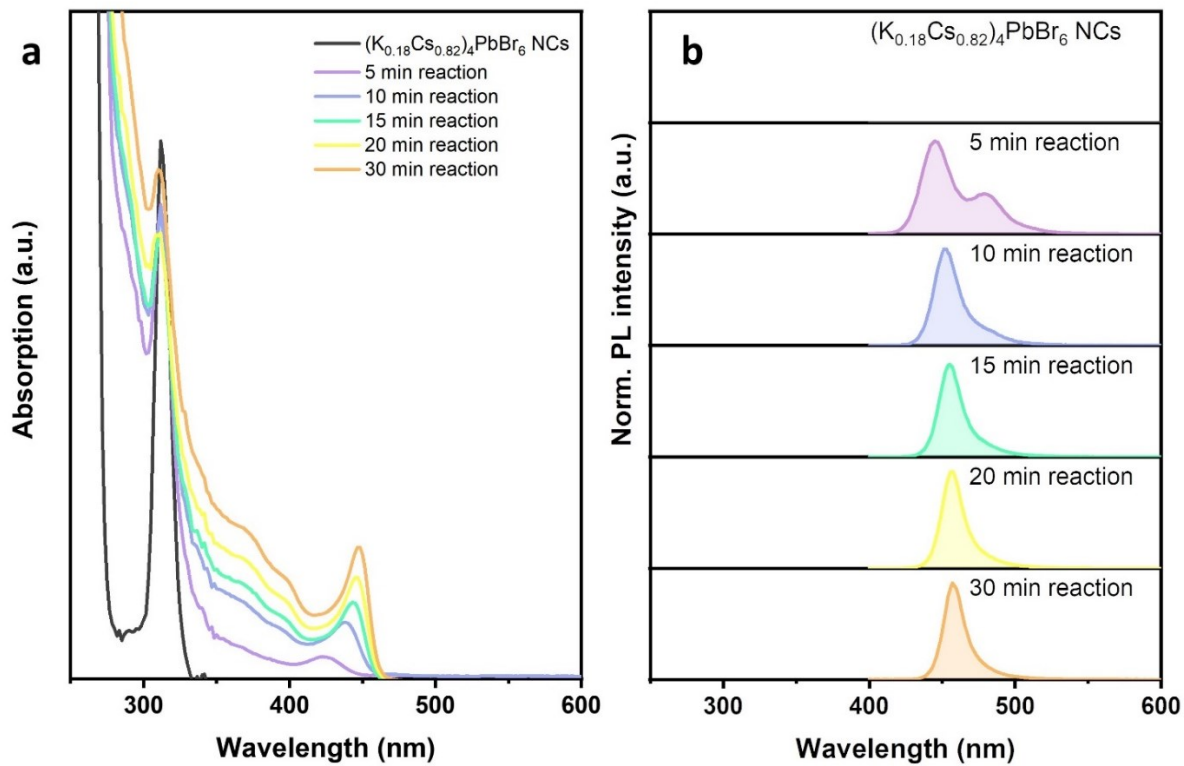

**Figure S16.** Absorption spectra evolution (a) and normalized PL spectra (b) of the CsPbBr$_3$ QDs at the early stages of the reaction.

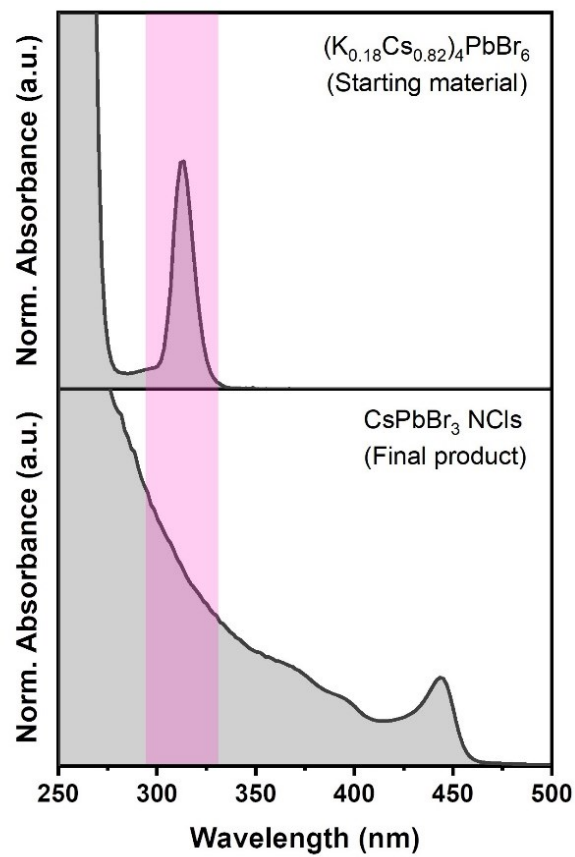

**Figure S17.** Normalized absorption spectra of the starting material (($K_{0.18}Cs_{0.82})_4PbBr_6$ nanocrystals) in the top panel and the final product ($CsPbBr_3$ QDs) in the bottom panel. The pink mark indicates the position of the excitonic peak of ($K_{0.18}Cs_{0.82})_4PbBr_6$ nanocrystals that we used as reference to follow the reaction and whose disappearance indicates the end of the reaction.

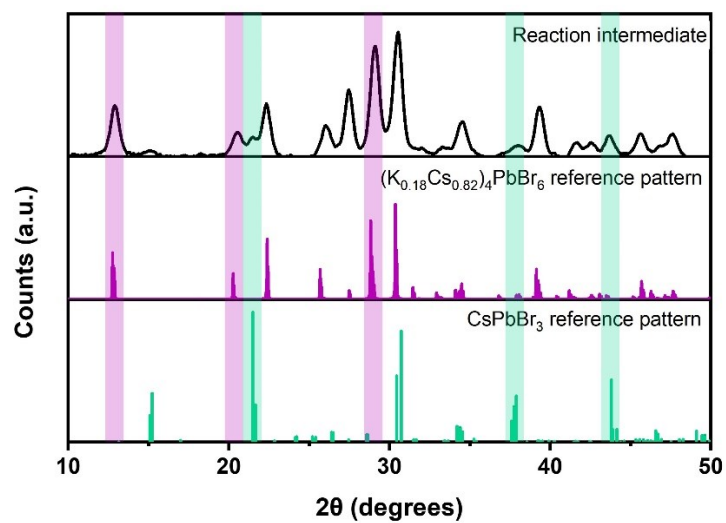

**Figure S18.** XRD diffraction pattern of the reaction intermediate during 0D phase $(K_{0.18}Cs_{0.82})_4PbBr_6$ nanocrystals transformation to 3D phase $CsPbBr_3$ QDs (black). The comparison with the corresponding reference patterns of $(K_{0.18}Cs_{0.82})_4PbBr_6$ (magenta) and $CsPbBr_3$ (green) indicates the coexistence of both phases during the reaction.

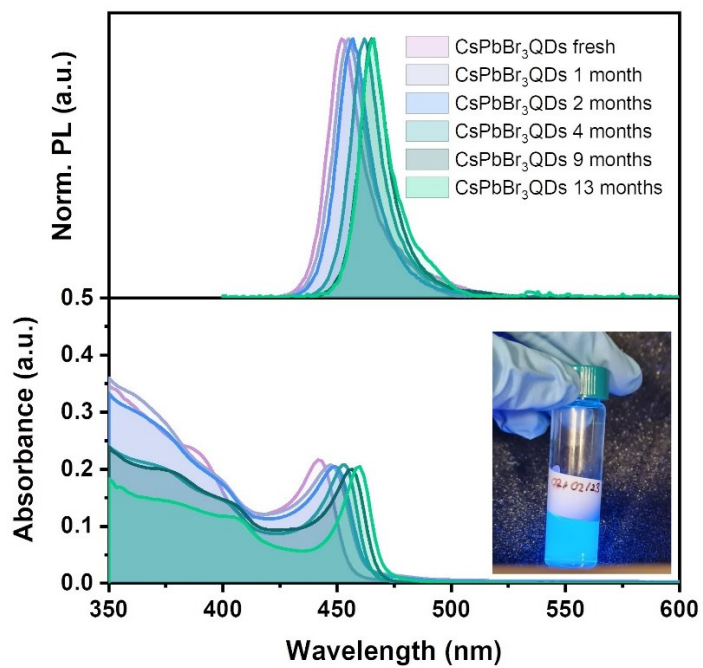

**Figure S19.** Absorption (bottom) and PL spectra of a colloidal solution of CsPbBr$_3$ QDs in toluene along 13 months. In the inset, photograph of the colloidal solution after 13 months.

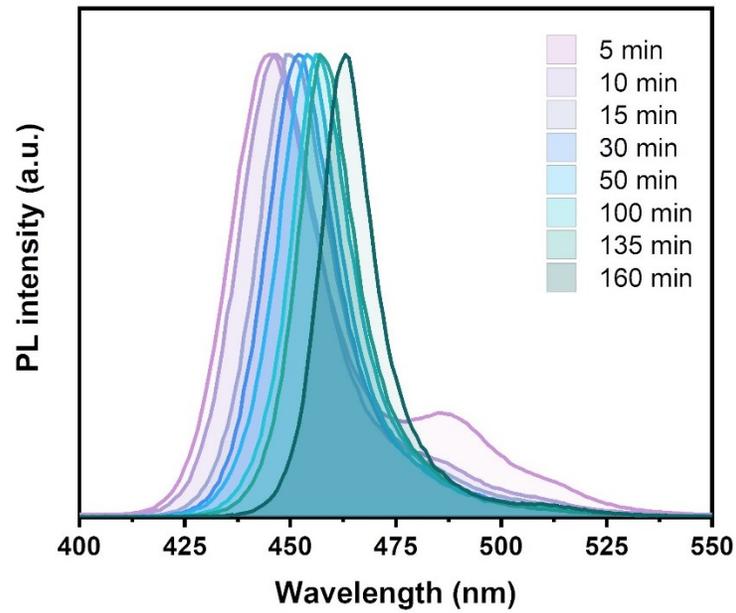

**Figure S20.** Normalized PL spectra of spectra of CsPbBr$_3$ QDs of different sizes obtained at different reaction times.

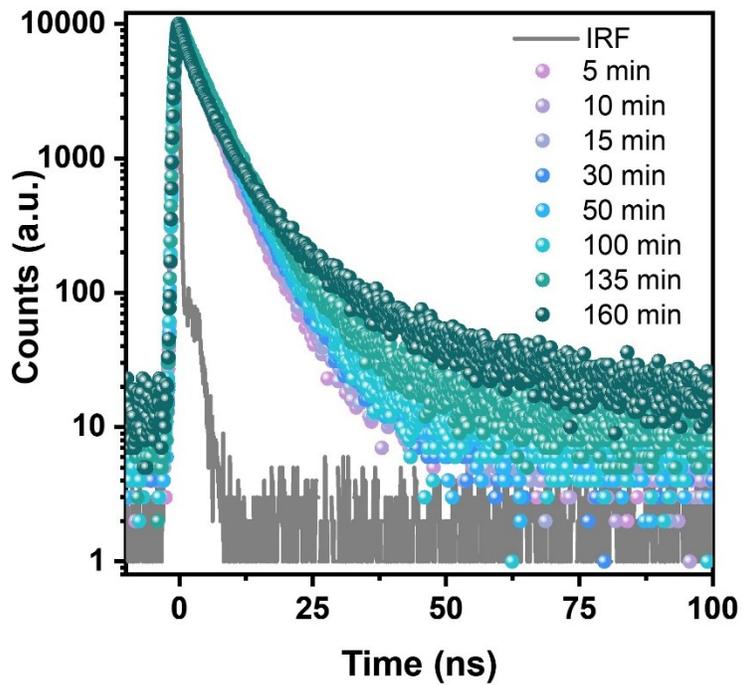

**Figure S21.** PL lifetime decays of CsPbBr$_3$ QDs synthesized at different reaction times obtained with a bi-exponential model.

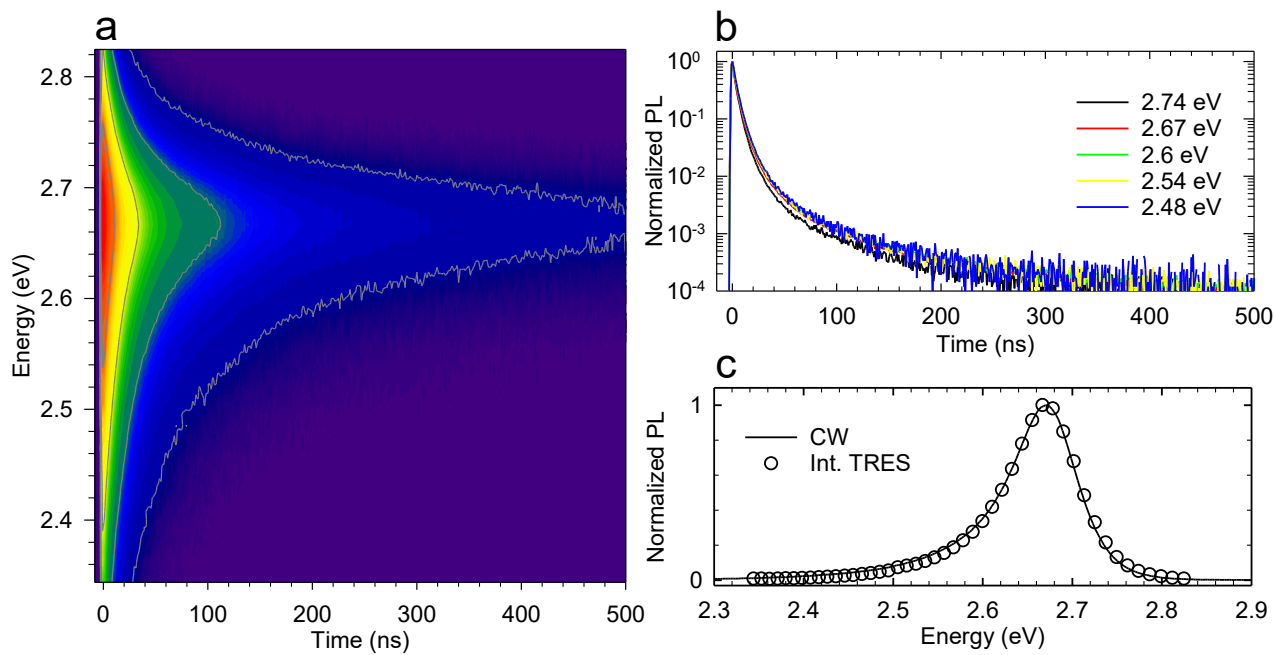

**Figure S22**. Contour plot of the spectrally resolved PL from CsPbBr$_3$ QDs (3.5 nm in size) under 3.05 eV excitation showing a uniform decay across the spectrum. The PL decay traces extracted at selected energies are shown in 'b'. c) Comparison between the time integrated PL spectrum and the cw PL spectrum showing perfect overlap.

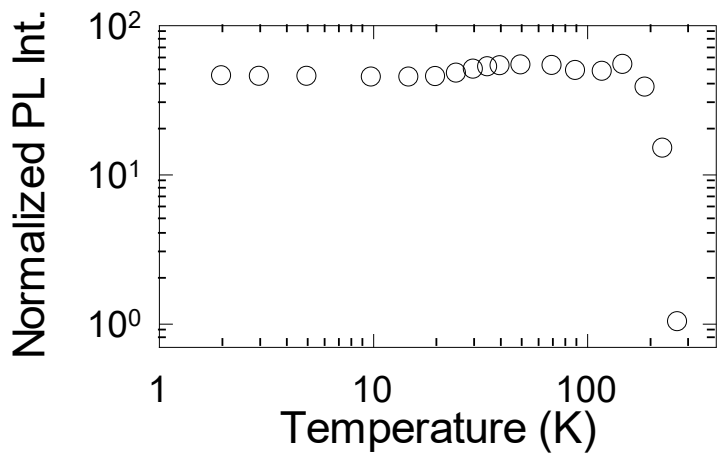

**Figure S23.** Temperature dependence of PL intensity for 3.5 nm CsPbBr$_3$ QDs. The data have been normalized for the value at room temperature.

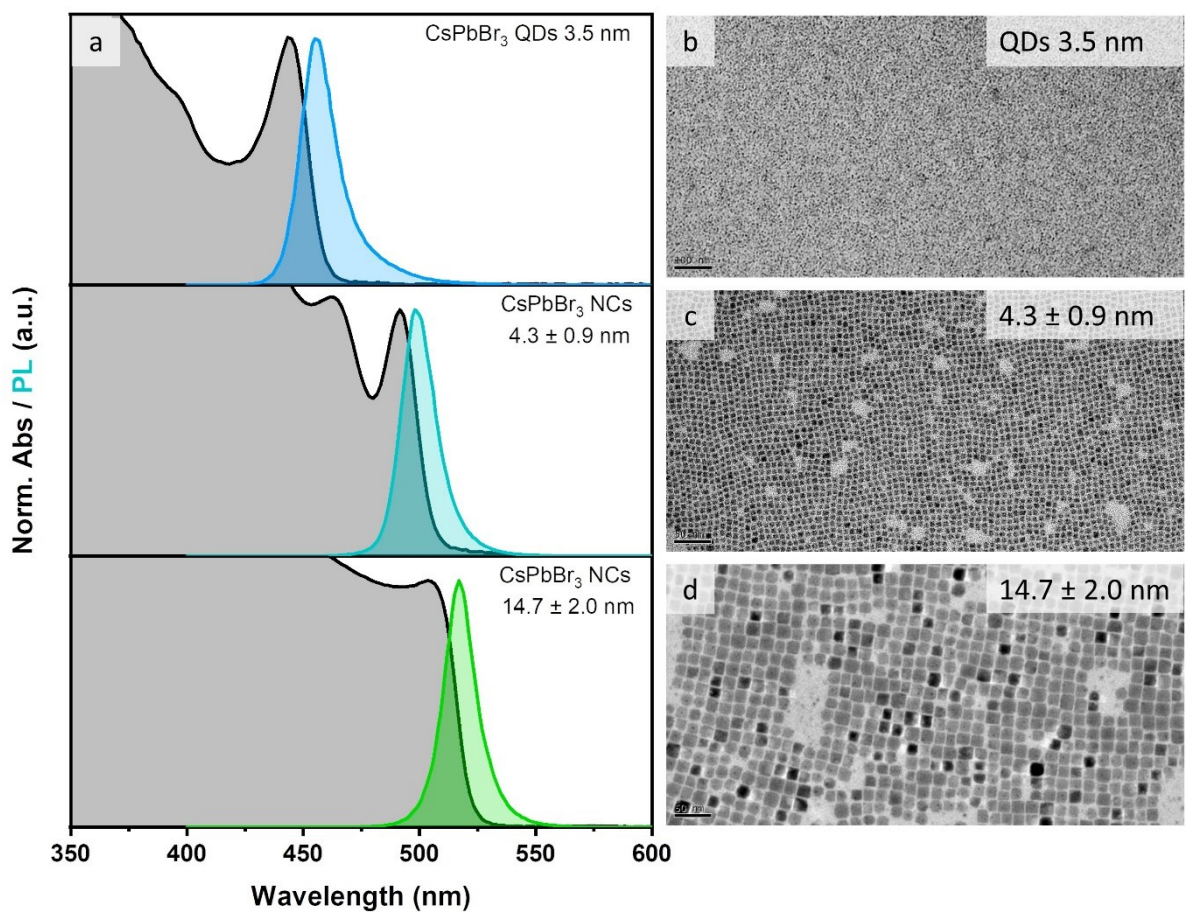

**Figure S24.** (a) Absorption and PL spectra of $CsPbBr_3$ QDs and nanocrystals with different lateral size. (b-d) Corresponding TEM images of 3.5 nm $CsPbBr_3$ QDs synthesized by $(K_{0.18}Cs_{0.82})_4PbBr_6$ recrystallization (b) and 4.3 nm $CsPbBr_3$ QDs (c) and 14.7 nm $CsPbBr_3$ nanocrystals (d), synthesized by a standard protocol previously reported.[2]